\documentclass[aps,prd,twocolumn,groupedaddress]{revtex4-1}

\usepackage{amsmath}
\usepackage{xspace}
\usepackage{xcolor}
\usepackage{amssymb}
\usepackage{graphicx}

\newcommand{\ket}[1]{\ensuremath{|#1\rangle}\xspace}
\newcommand{\bra}[1]{\ensuremath{\langle #1|}\xspace}

\newcommand{\dix}[1]{\cdot 10^{#1}}
\newcommand{\dif}[1]{\mathrm{d}#1\:} 
\newcommand{\diftrois}[1]{\mathrm{d}^3#1} 
\newcommand{\difv}[1]{\mathrm{d}^{3}\vect{#1}} 
\newcommand{\deriv}[2]{\frac{\mathrm{d}#1}{\mathrm{d}#2}} 
\newcommand{\derivdble}[3]{\frac{\mathrm{d}^2#1}{\mathrm{d}#2\mathrm{d}#3}} 
\newcommand{\pderiv}[2]{\frac{\partial #1}{\partial #2}} 
\newcommand{\pderivsec}[2]{\frac{\partial^{2} #1}{{\partial #2}^{2}}} 
\newcommand{\norm}[1]{\ensuremath{\left\Vert #1 \right\Vert}} 
\newcommand{\abs}[1]{\ensuremath{\left\vert #1 \right\vert}} 
\newcommand{\vect}[1]{\ensuremath{\vec{#1}}} 

\newcommand{\gdo}[1]{\bigcirc\left(#1\right)} 

\newcommand{\undemi}{{\frac{1}{2}}}

\begin{document}


\title{Quantum theory of curvature and synchro-curvature radiation in a strong and curved magnetic field, and applications to neutron star magnetospheres}


\author{Guillaume Voisin }
\email[]{guillaume.voisin@obspm.fr}
\author{Silvano Bonazzola}
\affiliation{LUTh, Observatoire de Paris, PSL Research University, 5 place Jules Janssen, 92190 Meudon, France}
\author{Fabrice Mottez}
\affiliation{LUTh, Observatoire de Paris, PSL Research University - CNRS,  5 place Jules Janssen, 92190 Meudon, France}


\date{\today}

\begin{abstract}
In a previous paper, we derived the quantum states of a Dirac particle in a circular, intense magnetic field in the limit of low momentum perpendicular to the field with the purpose of giving a quantum description of the trajectory of an electron, or a positron, in a typical pulsar or magnetar magnetosphere. 

Here we continue this work by computing the radiation resulting from transitions between these states. This leads to derive from first principles a quantum theory of the so-called curvature and synchro-curvature radiations relevant for rotating neutron-star magnetospheres.

We find that, within the approximation of an infinitely confined wave-function around the magnetic field and in the continuous energy-level limit, classical curvature radiation can be recovered in a fully consistent way. Further we introduce discrete transitions to account for the change of momentum perpendicular to the field and derive expressions for what we call quantum synchro-curvature radiation. Additionally, we express deconfinement and quantum recoil corrections. 
\end{abstract}


\maketitle

\section{Introduction}

In a previous paper \cite{voisin_curvature_1} (hereafter paper 1), we derived the states of an electron in a curved strong magnetic field within the approximation of a very low momentum perpendicular to the magnetic field. To ease calculations, it is convenient to consider a "circular" magnetic field, that is the field lines are of constant curvature and so form circles. In this paper, we compute the transition rates between these states in the limit of high momentum parallel to the magnetic field, in such a way that parallel transition can be considered approximately continuous. 
Our goal is to derive a quantum-electrodynamics theory of curvature radiation and low synchro-curvature radiation in the context of rotating-neutron-star magnetospheres. These magnetospheres are characterized by intense magnetic field from $10^5$ Teslas for recycled millisecond pulsars to $10^{11}$ Teslas for magnetars. The radius of curvature of magnetic field lines is typically larger than $10$ km, which is the typical radius of the star, within the assumption of a dipolar magnetic field.  Extremely large  electric-potential gaps along the open magnetic-field lines ( see e.g. \cite{arons_pulsar_2009} for a review) accelerate charged particles to energies only limited by radiation reaction at Lorentz factors as high as $10^5 - 10^8$. 

In this regime, an electron looses all of its momentum perpendicular to the magnetic field after traveling a few meters in the synchro-curvature regime (see hereafter and \cite{voisin_curvature_sf2a}). When only parallel momentum remains, radiation reaction is attributed to the so-called curvature radiation along the magnetic field \cite{ruderman_theory_1975}, which is the radiation of a charged particle following exactly a locally circular trajectory. Synchrotron radiation can be seen as a particular case where the trajectory is the cyclotron trajectory. However, curvature radiation usually refers to the case of a magnetic-field-line trajectory, and therefore is not strictly physical, in the sense that a particle not rotating around the field does not undergo any force capable of keeping it along. Therefore, curvature radiation is better seen as the mathematical zero-perpendicular-momentum limit of the so-called synchro-curvature radiation \cite{cheng_general_1996} that describes the classical theory of radiation by a charged particle with low perpendicular momentum in a locally circular magnetic field. Quantum corrections where added by \cite{zhang_quantum_1998} and latter by \cite{harko_unified_2002} in the form of an effective correction to classical expressions in analogy with equivalent photon theories developed  for synchrotron radiation which essentially amounts to the replacement $\omega \rightarrow \omega(1 + \hbar\omega/E)$ in the transition probability  $\omega^{-1}I(\omega)$ accounting or quantum recoil, where $I$ is the intensity per pulsation $\omega$ and $E$ is the energy of the particle. A formalism based on effective electric fields was developed  \cite{harko_unified_2002} to deal with further inhomogeneities of the magnetic field such as a perpendicular gradient of intensity.  A more compact but equivalent formalism for synchro-curvature radiation was also developed \cite{vigano_compact_2014}. Recently, a description  \cite{kelner_synchro-curvature_2015} with a self-consistent trajectory that takes carefully into account the drift along the cylinder generated by the circular field showed that the drift  effectively changes the curvature radius for relatively large Lorentz factors or low magnetic field intensities. 
As we pointed out in paper 1, classical synchro-curvature radiation results in numerous cases in a very fast decay of the perpendicular momentum of the particle which can reach the first Landau levels, if one assumes the well-known quantum theory of an electron in a homogeneous-intensity uniform-orientation magnetic field (see e.g. \cite{sokolovternov}). One then has to take into account discrete transitions from a Landau level to another \cite{latal_cyclotron_1986},  \cite{harding_quantized_1987}. This is particularly interesting when the plasma is at rest in the frame of the star such that the uniform-magnetic-field theory is locally relevant. This is the case for example in X-ray binaries where X-ray cyclotron lines have been observed and where two levels are typically separated by $11.6 B_8$ keV with $B_8 = B/(10^8\mathrm{Teslas})$  \cite{caballero_x-ray_2012}.

Therefore, classical synchro-curvature cannot hold for very low perpendicular momenta since the synchrotron part becomes discrete. This effect cannot be taken into account with the usual quantum recoil corrections which apply in the continuous limit. Besides it does not take into account the fact that two quantum numbers are changing, one for parallel and another for perpendicular momenta.  Simultaneously, cyclotron transitions are irrelevant for particles with high parallel momenta since they do not take into account longitudinal transitions, that is the curvature part of the radiation. In this paper, we start from first principles using the quantum states derived in paper 1. The resulting radiation results from transitions in the continuous approximation for parallel momentum variations and discrete for perpendicular momentum variations. Parallel transitions are treated in a similar way as \cite{sokolovternov} did for the quantum theory of synchrotron radiation (see also \cite{schwinger_quantum_1954} and \cite{schwinger_new_1978}). With this formalism syncho-curvature-like and curvature-like components appear in a very distinct fashion. As mentioned in paper 1, we neglect every drift of the particle, which is very appropriate except at extremes of the magnetic-field and Lorentz-factor ranges mentioned above. We also find additional corrections in $(\hbar\omega/E)^p (B_c/B)^q$ where $p,q$ are positive integers, $B_c = 4.4\dix{9}$ Teslas the critical field of Landau states and $B$ the magnetic-field intensity. We interpret these as deconfinement corrections, in the sense that they give the difference between a point-like particle and an extended wave-function around the magnetic-field line. At leading order, we find the classical curvature radiation. 

This paper is organized as follow: in section \ref{secradiationgeneral} we introduce the general formalism to compute quantum transitions, in section \ref{seccourbure} we develop this formalism in the particular case of curvature radiation which allows us to introduce notations and concepts that we generalize in section \ref{secgeneralcase} to the general case of synchro-curvature radiation, in section \ref{secpowerspectrum} we integrate the previously found expressions over solid angles to obtain power spectra and in section \ref{secconclusion} we discuss these results around the example of a millisecond pulsar.

\section{Radiation of a confined particles in quantum electrodynamics\label{secradiationgeneral}}

We compute the interaction of the electron with the photon vacuum to the first order of perturbation theory. The Hamiltonian of interaction is 
\begin{equation}
\hat{H}_{\text{int}} = \int  ec\overline{\Psi}_f\gamma^\mu\Psi_i  \hat{A}_\mu \diftrois{\vect{x}},
\end{equation}
where $\Psi_i$ is the initial state of the electron, $\overline{\Psi}_f = \Psi_f^*\gamma^0$ the Dirac conjugate of the final state. $\hat{A}$ is the vacuum amplitude operator (see e.g. \cite{bellac_physique_2003} equation 11.98), in the Heinsenberg representation 
\begin{equation}
\begin{array}{ll}
\hat{A}_\mu = \sqrt{\frac{\hbar}{2\epsilon_0 V}}\sum_{\vect{k}, e} \frac{1}{\sqrt{\omega_k}}\left( \right.& \left. a_{\vect{k}, z} e_\mu(\vect{k}) e^{\imath(\vect{k}\cdot\vect{x} - \omega_k t)} +  \right.\\
& \left. a_{\vect{k}, \epsilon}^\dagger e_\mu^*(\vect{k}) e^{-\imath(\vect{k}\cdot\vect{x} - \omega_k t)} \right)
\end{array},
\end{equation}
where we consider photons of four-vector $\left(\hbar\omega_k/c, \hbar\vect{k}\right) $ with polarizations $e_\mu(\vect{k}) = \left(e_0(\vect{k}), \vec{e}(\vect{k})\right)$ in the transverse (Coulomb) gauge such that: $\vect{k}\cdot \vect{e} =  0$. $\epsilon_0 \simeq 8.854\dix{-12} \text{ F}/\text{m}$ is the electric permittivity of vacuum and $V \equiv L^3$ the volume of quantification.

Since the number of electrons does not vary we need not quantify the electron field $\Psi$.

The rate of transition from vacuum to a state with one photon characterized by $(\vect{k}, e)$ while the electron switches from an initial state "i" to a final state "f" is given by 
\begin{equation}
 w_{fi} = \pderiv{}{t}\norm{\int_0^t\dif{\tau}e^{\imath\frac{E_f + \hbar\omega - E_i}{\hbar}\tau }\bra{1_{\vect{k}, e}, f} \frac{\hat{H}_\text{int}}{\hbar} \ket{0, i}}^2,
\end{equation}
which after standard manipulation ( e.g. \cite{berestetskii_quantum_1982}, \cite{sokolovternov}) gives
\begin{equation}
 w_{fi} = \norm{M_{fi}}^2 2\pi\hbar \delta\left(E_f + \hbar\omega - E_i \right) ,
\end{equation}
where $M_{fi} = \bra{1_{\vect{k}, \epsilon}, f} \frac{\hat{H}_\text{int}}{\hbar} \ket{0, i}$ is the matrix element of the transition, which in this case can be explicitly written for each mode as 
\begin{equation}
	\label{eqmatrixelement}
	M_{fi} =   e\frac{j^\mu e_\mu}{\sqrt{2\epsilon_0 \hbar \omega_k V}},
\end{equation}
where $j^\mu$ are the components of the transition current 
\begin{equation}
\label{eqcurrent}
	j^\mu = c\int \overline{\Psi}_f \gamma^\mu\Psi_i e^{-\imath\vect{k}\cdot\vect{x}}\diftrois{x}.
\end{equation}

In the continuum limit, we obtain the differential probability of radiating a photon in the solid angle $\dif{o}$ at a pulsation in $\dif{\omega}$ by multiplying by the density of such states $\frac{\omega^2\dif{o}\dif{\omega}}{c^3 (2\pi)^3 / V}$,
\begin{equation}
\label{eqdifprobaofemission}
	\dif{w_{fi}} = \norm{M_{fi}}^2 2\pi\hbar \delta\left(E_f + \hbar\omega - E_i \right) \frac{\omega^2\dif{o}\dif{\omega}}{c^3 (2\pi)^3 / V}.
\end{equation}

To obtain the radiated intensity we need only multiply by the photon energy $\hbar\omega$ the differential probability \eqref{eqdifprobaofemission}, and sum over every possible final energy states applying $\int \frac{\dif{E_f}}{\hbar\Omega}$ in the continuum limit and ultra-relativistic limit defined below along with $\Omega$. The intensity per pulsation per solid angle corresponding to a transition between an initial state $i$ and a final state $f$ reads 

\begin{equation}
\label{eqintensity}
 \derivdble{I_{f, i}^{\vect{e}}}{o}{\omega} = \frac{\hbar \omega^3 V}{\Omega(2\pi)^2 c^3} \norm{M_{fi}}^2(E_f = E_i -\hbar\omega).
\end{equation}

\section{Classical curvature radiation from quantum electrodynamics \label{seccourbure}}
\begin{figure}
\begin{center}
\includegraphics[width=0.45 \textwidth]{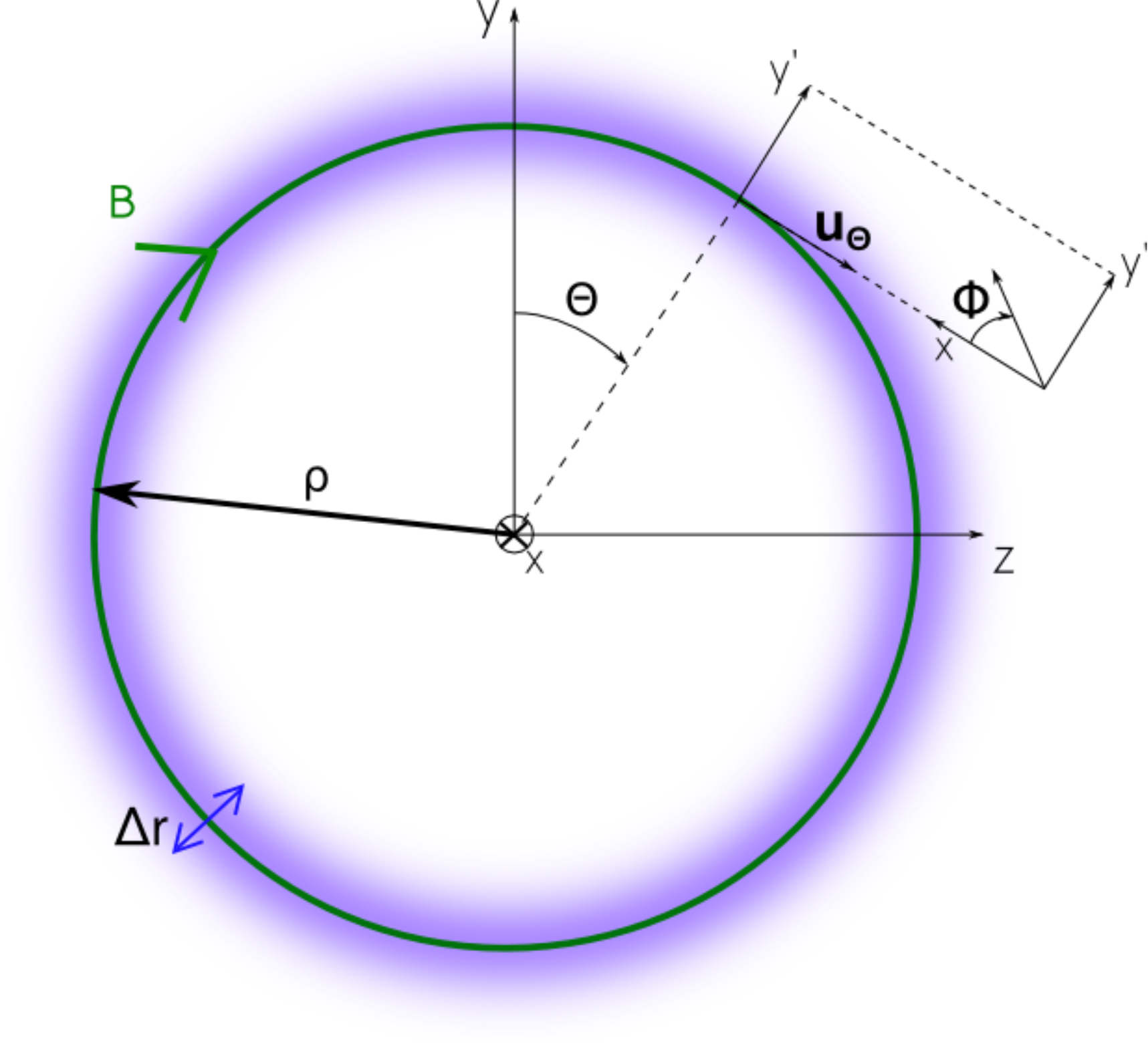}
\caption{\label{figcoord} Representation of a circular magnetic field line (green) of radius $\rho$, called "main circle" in the text. The blue shadow around the line represents the wave function of a ground orthogonal level with a characteristic extent $\lambda$. The relation between the toroidal coordinates $(r,\theta,\phi)$ and the cartesian coordinates $(x,y,z)$ is also shown. }
\end{center}
\end{figure}

In this paper we consider ultra-relativistic particles traveling along a circular magnetic field, the states of which were derived in paper 1 \cite{voisin_curvature_1}. The proper energies can be written 
\newcommand{\lpara}{{l_\parallel}}
\begin{equation}
\label{eqenergy}
E = \sqrt{m^2c^4 + 2m^2c^4\frac{B}{B_c} n + \hbar^2\Omega^2\lpara^2}
\end{equation}
where $B$ is the magnetic field,  $B_c = \frac{m^2c^2}{e\hbar} = 4.4\dix{9}$ Teslas is the critical magnetic field for which the difference between two Landau levels is equal to the rest mass energy of the electron, $\Omega = c / \rho$ is the pulsation of the particle along the main circle (see figure \ref{figcoord}). The numbers $n$ and $\lpara$ are integers respectively quantifying the angular momentum around the magnetic field and around the axis of the circular magnetic field (see figure \ref{figcoord}). 

In the theory of classical curvature radiation the rotation of the particle around the trajectory is neglected. Here we therefore take the lowest perpendicular state that is $n = 0$. Moreover, reminding that in the ultra-relativistic approximation most of the energy is in the longitudinal term, we expand the energy \eqref{eqenergy} as
\begin{equation}
\label{eqenergyUR1}
E = \hbar\Omega\lpara \left(1 + \frac{1}{2\gamma^2} +  \gdo{\frac{1}{\gamma^4}}\right)
\end{equation}
where $\gamma = E / (mc^2)$ 
is the classical Lorentz factor.  
The wave function  corresponding to this perpendicular fundamental state (see paper 1) is given to $\gdo{\gamma^{-2}}$ by
\begin{equation}
\Psi_0 = \frac{ e^{il_\parallel\theta}e^{-x^2/2}}{2\pi\sqrt{\rho \lambda^2}}\left(\begin{matrix}
i \sin\frac{\theta}{2}\\
-\cos\frac{\theta}{2} \\
-i\sin\frac{\theta}{2}\\
\cos\frac{\theta}{2} 
\end{matrix}\right),
\end{equation}
where $\rho$ is the radius of the classical trajectory that we call here the main circle and 
\begin{equation}
\label{eqmaglength}
\lambda = \left(\frac{2\hbar}{eB}\right)^{1/2}
\end{equation} is the magnetic length scale which characterizes the extent of the wave function perpendicular to the main circle. We use the toroidal coordinates related to the Cartesian system $(x,y,z)$ by the homeomorphism
\begin{equation}
\label{eqtorocoord}
 T: (r,\theta,\phi) \rightarrow \left(\begin{array}{l}
 x = r\cos\phi \\ 
 y = \cos\theta (\rho + r\sin\phi) \\ 
 z = \sin\theta(\rho + r\sin\phi)
 \end{array}
  \right),
\end{equation}
where $\theta$ represents the direct angle with respect to the $\vect{y}$ axis in the $(\vect{y}, \vect{z})$ plane,  $\phi$ represents the direct angle with respect to $\vect{x}$ in the plane $(\vect{x}, \vect{y'})$ of the local frame $(\vect{x}, \vect{y'},\vect{u}_\theta)$ image of $(\vect{x}, \vect{y},\vect{z})$ by a rotation of $\theta$ around $\vect{x}$ and $r$ represents the distance to the main circle. For further references on the coordinate system, see paper 1 and figure \ref{figcoord}. Here we use the reduced variable $x = r/\lambda$.  Moreover, the approximation used in paper 1 imposes that all our expressions are given to leading order in
\begin{equation}
\epsilon = \lambda / \rho \ll 1.
\end{equation}

We now have all the ingredients to compute the current \eqref{eqcurrent} for a transition between two perpendicular fundamentals of initial longitudinal number $\lpara_i$ and  final $\lpara_f$. It reads 
\begin{eqnarray}
\label{eqcurrent00}
j_{00} & = & \frac{1}{2\pi^2} \left(0,  \int   \sin\theta e^{-x^2}e^{i \left(l_i - l_f \right)\theta -i \lambda\vect{k}\cdot\vec{x}}\difv{x}, \right. \\
 & & \left. \int \cos\theta e^{-x^2}e^{i \left(l_i - l_f \right)\theta -i \lambda\vect{k}\cdot\vec{x}}\difv{x} \right) , 
\end{eqnarray}
with a dimensionless $\difv{x} =  x\dif{x}\dif{\theta}\dif{\phi} + \gdo{\epsilon}$. 

In the following we restrict ourselves to wave numbers laying in the $(\vect{z}, \vect{x})$ plane defined as
\begin{equation}
\label{eqk}
 \vect{k} = k(\sin\kappa,0,\cos\kappa)
\end{equation}
 where $\kappa$ is the direct angle from the $z$ axis. Since $\vec{x}$ is a symmetry axis, this is done without loss of generality. This allows us to choose the polarization basis (we use the same basis as in used in the textbook \cite{jackson_classical_1998})  
\begin{eqnarray}
\label{eqpolarizations}
 \vect{e}_\parallel & = & (0,1,0) , \nonumber \\ 
 \vect{e}_\bot & = & \frac{\vect{k}}{k}\wedge\vect{e}_\parallel = (-\cos\kappa, 0, \sin\kappa).
\end{eqnarray}
From a classical point of view, the parallel polarization $\vect{e}_\parallel$ points towards the center of the trajectory of the electron, and the perpendicular polarization $\vect{e}_\bot$ completes the direct triad $\left(\vec{k}/k, \vect{e}_\parallel, \vect{e}_\bot\right)$.

From equation \ref{eqenergyUR1}, one derives the relation between the variation of the parallel quantum number $\Delta\lpara = \lpara_i - \lpara_f$ and the variation of energy of the electron $E_i - E_f = \hbar\omega$, where $\omega$ is the pulsation of the emitted photon. Considering $\lpara$ as a continuous parameter, the energy variation can be Taylor expanded  
\begin{equation}
\label{eqdeltaE}
	\hbar\omega =   \Delta l \left.\pderiv{E_{l,\sigma}}{l}\right\vert_i - \frac{{\Delta l}^2}{2}\left.\pderivsec{E_{l,\sigma}}{l}\right\vert_i 
\end{equation}
\newcommand{\gammatild}{\tilde{\gamma}}
\newcommand{\ome}{ \frac{\hbar \omega}{E} }
\newcommand{\omec}{ \left( \frac{\hbar\omega}{E} \right)^{2} }
\newcommand{\bcsb}{\frac{B_c}{B}}
\newcommand{\bsbc}{\frac{B}{B_c}}
\newcommand{\Isin}{I_\mathrm{sin}}
\newcommand{\Icos}{I_\mathrm{cos}}
\newcommand{\omcrittilde}{\tilde{\omega}_\text{crit}}
\newcommand{\omcrit}{\omega_\mathrm{crit}}
\newcommand{\airy}{\mathrm{Ai}}
which is inverted into 
\begin{equation}
\label{eqdl00}
\Delta l= \frac{\omega}{\Omega}\left(1 + \frac{1}{2\gamma^2} \right) + \gdo{\frac{\hbar\omega}{E}}.
\end{equation}
We give additional $\hbar\omega/E$ terms, which are quantum recoil corrections, in the next sections. 


The imaginary exponential in the current \eqref{eqcurrent00} can be rewritten, using \eqref{eqdl00} and expanding the scalar product thanks to  \eqref{eqtorocoord} and \eqref{eqk}, as 
\begin{equation}
\label{eqphasetheta1}
e^{i\frac{\omega}{\Omega}\left(1 + \frac{1}{2\gamma^2}\right)\theta - i\rho k \cos\kappa\sin\theta}e^{-ix\lambda k\left(\cos\phi\sin\kappa + \sin\phi\cos\kappa\sin\theta\right)}.
\end{equation}
The second factor above exists only in the quantum mechanical theory. One can easily be convinced of that by noticing the presence of the magnetic length $\lambda$ \eqref{eqmaglength} which contains the Planck constant $\lambda \propto \hbar^{1/2}$. To obtain the classical theory one therefore puts $\lambda = \hbar = 0$. We neglect this factor (put it to 1) in the first part  of the following discussion and then reintroduce it. 

As in the usual treatment of classical synchrotron or curvature radiation (see e.g. \cite{jackson_classical_1998}) we consider the approximation of high frequency photons in which 
\begin{equation}
\label{eqhfapp}
\omega \gg \Omega.
\end{equation}
It follows that one can develop the phase in the first factor above to third order in $\theta$ since the exponential will oscillates heavily even for $\theta \ll 1$ as found in the literature on the classical radiation. One also expects a very high relativistic beaming implying that $\kappa \sim 1/\gamma$, and we can therefore expand $\cos \kappa = 1 - \frac{1}{2}\kappa^2 + \gdo{\kappa^4}$. We also notice that $\rho k = \omega / \Omega$. It follows that \eqref{eqphasetheta1} now reads
\begin{equation}
\label{eqphasetheta2}
e^{i\frac{\omega}{2\Omega}\left(\left(\kappa^2 + \frac{1}{\gamma^2}\right)\theta +\frac{\theta^3}{3}\right)}.
\end{equation}
We check the consistency of our approximations by looking at the qualitative behavior of \eqref{eqphasetheta2} above when integrated over $\theta$  as in \eqref{eqcurrent00}:
\begin{itemize}
\item When in the integral $\theta > \bar{\theta} = \left(\kappa^2 + \frac{1}{\gamma^2}\right)^{1/2} $ the $\theta^3$ term in the phase becomes dominant.
\item If $\frac{\omega}{\Omega}\bar{\theta}^3 > 1$ then the exponential oscillates heavily for $\theta \gg \bar{\theta}$ and kills the remaining part of the integral. This sets a critical pulsation $\omega \sim \bar{\theta}^{-3}\Omega$ above which the integral starts to decay.
\item The smallest critical pulsation corresponds to $\kappa = 0$. More generally, if $\kappa \gg 1/\gamma$ the transitions will remain possible on a much smaller part of the spectrum,and we recover the relativistic beaming condition that transitions are most likely for $\kappa \sim 1/\gamma$.  Further we use the definition given by, e.g, \cite{schwinger_classical_1949} or \cite{jackson_classical_1998} to define the critical pulsation of the dominant contribution as 
\begin{equation}
\label{eqomcritcourbure}
\omcrit = \frac{3}{2}\Omega\gamma^3.
\end{equation}
\item  As a result, the dominant contribution to the integral comes from the part where $\theta \sim 1/\gamma$. This justifies the earlier expansion of trigonometric functions in $\theta$. 
\end{itemize}

Let's reintroduce the second factor in \eqref{eqphasetheta1}.  If one assumes the previous result that $\theta \sim \kappa \sim 1/\gamma$ and $x \sim 1$ then the amplitude of the phase is about 
\begin{equation}
\label{eqphasescale}
\frac{\lambda k}{\gamma} = \frac{\hbar\omega}{E}\left(\frac{2B_c}{B}\right)^{1/2}. 
\end{equation}
Therefore there is a range of magnetic fields and electron energies (remember that $\omega \sim \omcrit$) for which this amplitude is small. For example, for a "typical" pulsar with $B = 10^8$ Teslas, $\gamma = 10^7$ and a dipolar magnetic field with curvature next to the pole of $\rho = 10^4$ m (see e.g. \citep{arons_pulsar_2009}) one has 
\begin{equation}
\frac{\lambda \omcrit/c}{\gamma} \simeq 0.05 \gamma_7^{2} \rho_4^{-1} B_8^{-1/2}.
\end{equation} 
For now, we can legitimately consider these corrections to be negligible. This amounts to consider that the particle is infinitely confined, $\lambda = 0$, as in the classical theory. We bring back the deconfinement corrections in the next sections. 

We proceed to integrate the expressions in the current \eqref{eqcurrent00}. Integration over $\phi$ simply yields a factor $2\pi$ since within our approximation of infinite confinement there is no explicit dependence in $\phi$. Integration over $x$ of $xe^{-x^2}$ yields a factor $1/2$.  To integrate over $\theta$, we use the fact that $\sin\theta$ and $\cos\theta$ are slowly varying compared to the exponential for $\theta \gg 1/\gamma$ to develop them to first order in $\theta$. Moreover, we extent the boundaries to infinity since the contributing part is centered on $\theta \ll 1$. We get the two following integrals 

\begin{eqnarray}
\label{eqI1_2}
I_\mathrm{cos} & = &  \int_{-\infty}^{+\infty} e^{i \frac{\omega}{2\Omega}\left(\theta\left(\kappa^2 + \frac{1}{ \gammatild}\right) + \frac{\theta^3}{3} \right)} \dif{\theta}, \\
\label{eqIsin_2}
I_\mathrm{sin} & = & \int_{-\infty}^{+\infty} \theta e^{i \frac{\omega}{2\Omega}\left(\theta\left(\kappa^2 + \frac{1}{ \gammatild}\right) + \frac{\theta^3}{3} \right)} \dif{\theta},\\
\end{eqnarray}
and
\begin{equation}
j_{00} = \frac{1}{2\pi}\left(0, \Icos, \Isin \right).
\end{equation}

We recognize in \eqref{eqI1_2} an Airy integral and its derivative in \eqref{eqIsin_2}. We use in this paper the definitions of special functions of \cite{olver_nist_2010} where the Airy function is given by
\begin{equation}
\label{eqAidef}
\airy(x) = \frac{1}{2\pi}\int_{-\infty}^\infty \dif{t} e^{i\left(xt + \frac{t^3}{3}\right)}.
\end{equation}
After performing the change of variable 
\begin{equation}
\label{eqvarchangeairy}
\tilde{\theta} \rightarrow \theta = \left(\frac{\omega}{2\Omega}\right)^{-1/3}\tilde{\theta}
\end{equation}
one identifies $x = (\omega/2\Omega)^{2/3}\left(\kappa^2 + \frac{1}{\gammatild^2}\right) $ and obtains
\begin{eqnarray}
\label{eqI1_3}
I_\mathrm{cos} & = & 2\pi (\omega/2\Omega)^{-1/3} \airy(x), \\
\label{eqIsin_3}
I_\mathrm{sin} & = & -2\pi  (\omega/2\Omega)^{-2/3} \airy'(x). \\
\end{eqnarray}
For practical calculations, the Airy integrals can be changed into modified Bessel functions $K_\nu$
\begin{eqnarray}
K_{1/3}(\xi)& =& \pi\sqrt{\frac{3}{\abs{x}}}\airy(x), \\
K_{2/3}(\xi)& =& -\pi\frac{\sqrt{3}}{x}\airy'(x),
\end{eqnarray}
with $\xi = \frac{2}{3}\abs{x}^{3/2}$ and assuming $x > 0$ .
 
We now calculate the intensities. We need to compute the matrix elements \eqref{eqmatrixelement} for both parallel and perpendicular polarizations. We seek a result to the lowest ultra-relativistic order. For this, it is useful to see that owing to the $\theta$ factor in \eqref{eqIsin_2} $\Isin \sim \frac{1}{\gamma}\Icos$. Further, the polarization vectors \eqref{eqpolarizations} are expanded to first order in $\kappa \sim 1/\gamma$ such that the squared matrix elements for respectively parallel and perpendicular polarizations are
\begin{eqnarray}
{M_{00}^{\parallel}}^2 & = & \frac{e^2}{2\epsilon_0 \hbar \omega_k V} \Isin^2, \\
{M_{00}^{\perp}}^2 & = & \frac{e^2}{2\epsilon_0 \hbar \omega_k V} \kappa^2\Icos^2. 
\end{eqnarray}
 Inserting the above matrix elements in the expression of the intensity \eqref{eqintensity} and expressing $\Icos$ and $\Isin$ with modified Bessel functions one obtains 

\begin{eqnarray}
\label{eqclasscurvature1}
 \derivdble{I_{00}^{\parallel}}{o}{\omega} & =  &\frac{1}{2\pi\Omega}\frac{e^2\omega^2}{12\pi^3\epsilon_0 c} \left( \kappa^2 + \frac{1}{\gamma^2}\right)^2K_{2/3}^2(\xi) ,\\
 \label{eqclasscurvature2}
   \derivdble{I_{00}^{\bot}}{o}{\omega} & =  & \frac{1}{2\pi\Omega}\frac{e^2\omega^2}{12\pi^3\epsilon_0 c} \kappa^2\left( \kappa^2 + \frac{1}{\gamma^2}\right)K_{1/3}^2(\xi),
\end{eqnarray}
where $\xi  = \frac{\omega}{3\Omega}\abs{\kappa^2 + \frac{1}{\gamma^2}}^{3/2}$ .
These expressions are identical to expressions found in the classical theory (see e.g. \cite{jackson_classical_1998}.

\section{General calculation of synchro-curvature including quantum corrections \label{secgeneralcase}}
We now generalize the calculation of the previous section to transitions between states of any initial perpendicular quantum number $n$ to a final number $n'$ including quantum corrections up to second order in $\frac{\hbar\omega}{E}$. The need to go to second order is dictated by the occurence of deconfinement corrections in $B_c/B$  potentially increasing the role of this order for relatively low magnetic fields, as we see in \eqref{eqphasescale2}. 

The energy of an ultra-relativistic particle of perpendicular quantum number $n$ is generalized from \eqref{eqenergy} and \eqref{eqenergyUR1} as
\begin{equation}
\label{eqenergyURn}
E = \hbar\Omega\lpara \left(1 + \frac{1}{2\gamma^2} +  \frac{1}{\gamma^2}\frac{B}{B_c}n + \gdo{\frac{1}{\gamma^4}}\right).
\end{equation}
For $n > 0$, the perpendicular quantum number is degenerate between the perpendicular angular momentum $l_\perp$ and the center-of-trajectory quantum number $s$ since $n = l_\perp + s$ (see paper 1). Without loss of generality, we can consider only centered trajectories with $s=0$. To energies \eqref{eqenergyURn} then correspond the proper states found in paper 1 with $n = l_\perp$ that we develop here to first ultra-relativistic order in $1/\gamma$,
\begin{widetext}
\begin{equation}
\label{eqpsin}
\Psi_n(x,\theta, \phi) = \frac{e^{\frac{-x^2}{2}}x^{n-1}e^{i(n-1)\phi}}{2\pi\sqrt{\Gamma(1+n)\rho\lambda^2}}\left(
\begin{array}{l}
\zeta i x e^{i\phi}\sin\frac{\theta}{2} + \frac{i}{\gamma}\left(\frac{1}{2}\zeta x e^{i\phi}\sin\frac{\theta}{2} - \frac{\zeta -1}{2}n\sqrt{2}\left(\bsbc\right)^{1/2}\cos\frac{\theta}{2}\right)\\
-\zeta x e^{i\phi} \cos\frac{\theta}{2} -\frac{1}{\gamma}\left(\frac{1}{2}\zeta x e^{i\phi}\cos\frac{\theta}{2} + \frac{\zeta -1 }{2}n\sqrt{2}\left(\bsbc\right)^{1/2}\sin\frac{\theta}{2}\right) \\
-i \zeta x e^{i\phi}\sin\frac{\theta}{2} + \frac{i}{\gamma}\left(\frac{1}{2}\zeta x e^{i\phi}\sin\frac{\theta}{2} + \frac{1+\zeta}{2}n\sqrt{2}\left(\bsbc\right)^{1/2}\cos\frac{\theta}{2}\right) \\
\zeta x e^{i\phi}\cos\frac{\theta}{2} - \frac{1}{\gamma}\left(\frac{1}{2}\zeta x e^{i\phi}\cos\frac{\theta}{2} - \frac{1+\zeta}{2}n\sqrt{2}\left(\frac{B}{B_c}\right)^{1/2}\sin\frac{\theta}{2}\right)
\end{array}\right).
\end{equation}
\end{widetext}
The parameter $\zeta =\pm 1$ describes the spin orientation and is degenerate with respect to the energy. 

We now outline the computation from the transition currents $j_{nn'}$ to the intensities. We assume $n > n'$ without loss of generality. Putting \eqref{eqpsin} in the current \eqref{eqcurrent} and projecting onto polarizations \eqref{eqpolarizations} one obtains the following structure 
\begin{equation}
\label{eqjnnp}
j_{nn'}^{\mu}e_{\mu}^{\sigma} = \zeta\zeta'\int\diftrois{x}\sum_{p = n - 1 - n'}^{n + 1 - n'} a_p e^{ip\phi}e^{i \left(l_i - l_f \right)\theta -i \lambda\vect{k}\cdot\vec{x}}
\end{equation}
where $\sigma$ denotes parallel or perpendicular polarization 	and each $a_p$ coefficient is of the form 
\begin{equation}
C(\kappa)x^{m_1}e^{-x^2}\cos^{m_2}\theta \sin^{m_3}\theta ,
\end{equation}
where $C$ is a coefficient depending only on $\kappa$ and $m_1, m_2, m_3$ are positive integers. 

In this section we take into account corrections to second order in $\ome$ which leads to express the variation of the quantum number $\lpara$ as
\begin{eqnarray}
\Delta l & = & \frac{\omega}{\Omega}\left\{1 + \frac{1}{2\gamma^2}\left[\left(1 + 2n\frac{B}{B_c} \right) \left(1 + \ome \right) \right. \right.\\ 
 & & \left.\left.-   2\frac{E}{\hbar\omega}\bsbc\Delta n\right] \right\}  + \gdo{\left(\ome\right)^{3}} \nonumber
\end{eqnarray}
where $\Delta n = n - n'$ . 
\newcommand{\gammadeux}{\gamma_2}
We see that the rightmost  exponential factor in \eqref{eqjnnp} takes exactly the same form as in \eqref{eqphasetheta1} if we make the replacement $\frac{1}{\gamma^2 } \rightarrow \frac{1}{\gammadeux}$ where we define 
\begin{equation}
\label{eqgammadeux}
\frac{1}{\gammadeux} = \frac{1}{\gamma^2}\left[\left(1 + 2n\frac{B}{B_c} \right) \left(1 + \ome \right) -   2\frac{E}{\hbar\omega}\bsbc\Delta n\right]
\end{equation}

Let's detail this effective Lorentz factor a little. The left part corresponds to transitions where the particle remains on the same perpendicular level $n$, with $(1+\hbar\omega/E)$ giving the high-energy quantum recoil correction. If $n=0$ and we neglect the high-energy correction we therefore recover $1/ \gamma^2$ as in the previous section. The second term results from the shift to a different perpendicular level. This term is particularly important for low-energy photons and high magnetic fields. Notice that it can even lead to a negative $\gamma_2$ meaning that energy is transferred from the perpendicular excitation of the electron to its longitudinal motion. We can follow the same reasoning as in previous section (see \eqref{eqphasetheta2} and thereafter) and obtain similar scalings provided one makes the replacement $\gamma \rightarrow \gammatild$ with 
\begin{equation}
\label{eqgammatild}
\gammatild = \sqrt{\abs{\gammadeux}},
\end{equation}
then 
\begin{equation}
\kappa\sim \theta \sim 1/\gammatild,
\end{equation}
and the critical pulsation 
\begin{equation}
\omcrittilde = \frac{3}{2}\Omega\gammatild^{3}. 
\end{equation}

We now proceed to integrate over $\phi$. To obtain the relevant high-energy accuracy to second order one separates the imaginary exponential in \eqref{eqjnnp} as in \eqref{eqphasetheta1} and notices that, similarly to \eqref{eqphasescale}, its argument is of order 
\begin{equation}
\label{eqphasescale2}
x\frac{\lambda k}{\gammatild} \sim \sqrt{n}\frac{\gamma}{\gammatild}\frac{\hbar\omega}{E}\left(\frac{2B_c}{B}\right)^{1/2}, 
\end{equation} 
where we used the fact that the averaged normalized radial distance of an electron is $\sim\sqrt{n}$ as explained in paper 1. Assuming \eqref{eqphasescale2} is small  compared to one, we expand the second factor of \eqref{eqphasetheta1} to second order in the argument $-ix\lambda k\left(\cos\phi\sin\kappa + \sin\phi\cos\kappa\sin\theta\right)$. We  are left to integrate terms of the form
\begin{eqnarray}
A_{pq} & = &\int_{-\pi}^{\pi}\dif{\phi} e^{ip\phi}\left(a\cos\phi + b\sin\phi\right)^q  
\end{eqnarray}
where $p$ and $q$ are integers and $q \geq 0$, $a$ and $b$ can have any value independent of $\phi$. One can show that 
\begin{eqnarray}
A_{pq} = 0 & \:\: \mathrm{ if} \:\: & \left\{ \begin{array}{l}
 q < \abs{p} \\
 \mathrm{or} \\
 q-\abs{p} \: \mathrm{odd}
 \end{array}\right. .
\end{eqnarray}
For this reason, the only transitions  yielding terms of order lower or equal to $\omec$ once current \eqref{eqcurrent} is inserted in the squared matrix element \eqref{eqmatrixelement} are for $n = n'$ ,$n' = n -1$ and $n' = n -2$. Moreover, one can see that the next non-null term of the expansion of \eqref{eqphasetheta2} is of order $\left(\ome\right)^{4}$, pushing further the validity of our approximation. In practice, we need 
\begin{eqnarray}
A_{pq} & = &\left\{ \begin{array}{lll}
\pi(a +ib) &  p=-1 & q = 1\\
\pi(a^2 + b^2) & p= 0 & q = 2 \\
\pi(a - ib) & p = 1 & q = 1 \\
\frac{\pi}{2}(a + ib)^2 & p = 2 & q = 2 
\end{array}\right. . 
\end{eqnarray}

We then integrate over $x$ with only integrals of the type
\begin{equation}
\int_0^{\infty}\dif{x}e^{-x^2} x^{2p+1} = \frac{p!}{2}, 
\end{equation}
where $p$ is a positive integer. 

We are now left with integrals over $\theta$ of the type 
\begin{equation}
B_{pq} = \int_{-\pi}^{\pi}\dif{\theta}e^{i\frac{\omega}{2\Omega}\left(\left(\kappa^2 + \frac{1}{\gammadeux}\right)\theta +\frac{\theta^3}{3}\right)}\cos^{p}\theta \sin^{q}\theta
\end{equation}
where $p,q$ are positive integers.
As in the previous section, the smallness of contributing values of $\theta \sim 1/\gammatild$ allows to extend boundaries to infinity. Moreover, to leading ultra-relativistic order one has 
\begin{equation}
\forall (p,q), B_{pq} =  \int_{-\infty}^{\infty}\dif{\theta}e^{i\frac{\omega}{2\Omega}\left(\left(\kappa^2 + \frac{1}{\gammadeux}\right)\theta +\frac{\theta^3}{3}\right)}\theta^q .
\end{equation}
Using definition \eqref{eqAidef} one sees that $B_{pq}$ is proportional to the q-th derivative of the Airy function.  Reminding that the Airy function verifies the relation \cite{olver_nist_2010}
\begin{equation}
\airy''(x) = x \airy(x)
\end{equation}
one is able to express every $B_{pq}$ in terms of $\airy$ and $\airy'$, and from that in terms of $\Icos$ and $\Isin$ (\eqref{eqI1_2},\eqref{eqIsin_2}). In particular, we need the following expressions 
\begin{eqnarray}
B_{p2} & = & -\left(\kappa^2 + \frac{1}{\gammadeux}\right)\Icos, \\
B_{p3} & = & \frac{1}{\gammatild^3}\frac{4\omcrittilde}{3\omega}\Icos - \left(\kappa^2 + \frac{1}{\gammadeux}\right)\Isin, \\
\end{eqnarray}
where the replacement $1/\gamma^2\rightarrow 1/\gammadeux$ is assumed in $\Icos$ and $\Isin$.

Squarring \eqref{eqjnnp}, inserting it into the matrix element \eqref{eqmatrixelement} and using formula \eqref{eqintensity} we obtain all the relevant intensities to order $\omec$. These intensities are proportional to $(\zeta\zeta')^2$ and therefore the spin average $\undemi \sum_{\zeta,\zeta' = \pm 1}$ is immediate, giving

\begin{widetext}
\begin{eqnarray}
\label{eqIbase1}
\derivdble{I_{nn}^{\parallel}}{o}{\omega}  & = & \frac{1}{2\pi\Omega}\frac{e^2\omega^2}{16\pi^3\epsilon_0 c} \left[\Isin^2 + \bcsb\omec\frac{\gamma^2}{\gammadeux}(n+1)\left( \Isin^2 - \frac{4}{3}  \frac{\omcrittilde}{\omega}\frac{\Icos}{\gammatild}\Isin \right) +  \omec n^2 \kappa^2 \Icos^2\right], \\
\label{eqIbase2}
\derivdble{I_{nn}^{\perp}}{o}{\omega} &  = &  \frac{1}{2\pi\Omega}\frac{e^2\omega^2}{16\pi^3\epsilon_0 c} \left[\kappa^2\Icos^2 + \bcsb\omec\frac{\gamma^2}{\gammadeux}(n+1)\kappa^2\Icos^2 + \omec n^2 \Isin^2 \right], \\
\derivdble{I_{nn-1}^{\parallel}}{o}{\omega} & = &  
 \frac{1}{2\pi\Omega}\frac{e^2\omega^2}{16\pi^3\epsilon_0 c} \left[\frac{B}{B_c}\frac{n}{2}\frac{\Icos^2}{\gamma^2} + \ome n \left(\frac{1}{\gammadeux} + \kappa^2\right)\Icos^2 + \bcsb\omec\frac{n}{2}\left(\gamma^2\kappa^2\Isin^2 + \gamma^2\left(\frac{1}{\gammadeux} + \kappa^2 \right)^2\Icos^2\right) + \right. \nonumber\\
& & \left. \omec\frac{n^2}{4}\left(\frac{2}{\gammadeux} - \kappa^2\frac{n-1}{n}\right)\Icos^2 \right], \\
\derivdble{I_{nn-1}^{\perp}}{o}{\omega} & = &  \frac{1}{2\pi\Omega}\frac{e^2\omega^2}{16\pi^3\epsilon_0 c} \left[\frac{B}{B_c}\frac{n}{2}\frac{\Icos^2}{\gamma^2} - \ome n \kappa^2\Icos^2 + \bcsb\omec\frac{n}{2}\gamma^2\kappa^2\left(\Isin^2 + \kappa^2 \Icos^2\right) + \right. \nonumber \\
 & & \left. \omec\frac{n^2}{4}\left(\frac{2}{\gammadeux} + \frac{n-1}{n}\kappa^2\right) \Icos^2\right], \\
\derivdble{I_{nn-2}^{\parallel}}{o}{\omega} & = &  \frac{1}{2\pi\Omega}\frac{e^2\omega^2}{16\pi^3\epsilon_0 c} \left[\omec\frac{n(n-1)}{4}\left(\Isin^2 + \kappa^2\Icos^2\right)\right], \\
\label{eqIbase4}
\derivdble{I_{nn-2}^{\perp}}{o}{\omega} & = &  \frac{1}{2\pi\Omega}\frac{e^2\omega^2}{16\pi^3\epsilon_0 c} \left[\omec\frac{n(n-1)}{4}\Isin^2\right].
\end{eqnarray}
\end{widetext}
Our result is based on the following hierarchy of scales 
\begin{equation}
\label{eqapproxs}
\frac{1}{\gamma} \ll \sqrt{n}\frac{\gamma}{\gammatild}\frac{\hbar\omega}{E}\left(\frac{2B_c}{B}\right)^{1/2} < 1 \:\mathrm{and} \: \frac{1}{\gamma}  \ll \bsbc .
\end{equation}
This allows to consider that all the gamma parameters have roughly the same order of magnitude compared to other terms $ 1/\gammatild^2 \sim 1/\gammadeux \sim 1/\gamma^2$. 
All terms are of second ultra-relativistic order since $\Isin \sim \Icos/\gammatild$ and $\kappa^2 \sim 1/\gammatild^2 $. One notices that this is not a strict expansion in powers of $\ome$ and $\bsbc$, since $\gammatild$ also contains such terms. It would even be impossible to perform a total, rapidly converging expansion of $\Isin, \Icos$ with respect to  $\bsbc$ since it is not necessarily small. However, the present expansion is relatively compact and directly reflects the confinement corrections as explained in \eqref{eqphasescale} and \eqref{eqphasescale2}. 

One recognizes the classical curvature intensities derived in the previous section, \eqref{eqclasscurvature1} and \eqref{eqclasscurvature2}, as the first terms of \eqref{eqIbase1} and \eqref{eqIbase2} respectively.

\section{Power spectrum \label{secpowerspectrum}}
We proceed to integrate expressions \eqref{eqIbase1}-\eqref{eqIbase4} over the solid angle $\dif{o}$ which can be explicited as
\begin{equation}
\dif{o} = \cos\kappa \dif{\kappa}\dif{\chi}
\end{equation}
where $\chi$ is an angle around the main circle. Integration of $\chi$ is trivial and yields a factor of $2\pi$. Integration over $\kappa$ requires more care. Applying the change of variable \eqref{eqvarchangeairy} we express all the relevant integrals over $\kappa$ of \eqref{eqIbase1}-\eqref{eqIbase4} in terms of the integrals calculated in appendix \ref{apsquaredairy} $I_a(\xi)$, $I_b(\xi)$, $I_c(\xi)$, $I_d(\xi), I_e(\xi)$ and  $I_f(\xi)$
\begin{eqnarray}
\label{eqrep1}
\int_{-\infty}^{\infty}I_{\mathrm{sin}}^2 \dif{\kappa} & = & \frac{\pi}{\sqrt{3}\gammatild^2}\frac{2\Omega}{\omega} I_a(\xi),\\
\int_{-\infty}^{\infty}I_{\mathrm{cos}}^2 \dif{\kappa} & = & \frac{2\pi}{\sqrt{3}}\frac{2\Omega}{\omega} I_b(\xi),\\
\int_{-\infty}^{\infty}\kappa^2 I_{\mathrm{cos}}^2 \dif{\kappa} & = & \frac{\pi}{\sqrt{3}\gammatild^2}\frac{2\Omega}{\omega}I_c(\xi), \\
\int_{-\infty}^{\infty} \Icos\Isin \dif{\kappa} & = & \frac{\pi}{\sqrt{3}\gammatild}\frac{2\Omega}{\omega}I_d(\xi), \\
\int_{-\infty}^{\infty}\kappa^2 \Isin^2 \dif{\kappa} & = & \frac{\pi}{4\sqrt{3}\gammatild^4}\frac{2\Omega}{\omega}I_e(\xi),\\
\label{eqrep6}
\int_{-\infty}^{\infty}\kappa^4 \Icos^2 \dif{\kappa} & = & \frac{\pi\sqrt{3}}{4\gammatild^4}\frac{2\Omega}{\omega}I_f(\xi). 
\end{eqnarray}
where we define 
\begin{equation}
\xi = \frac{\omega}{\omcrittilde}.
\end{equation}

The values of the previous integrals are summarized here by
\begin{eqnarray}
\label{eqIa}
I_a(\xi) & = & \left\{\begin{array}{ll}
\int_\xi^\infty K_{5/3}(x)\dif{x}   + K_{2/3}(\xi) & \gammadeux > 0\\
\pi\sqrt{3} -  \int_\xi^\infty \dif{x}F_{1/3}(x,\gammadeux) -  & \gammadeux < 0 \\
3F_{2/3}(\xi,\gammadeux) 
\end{array}\right. ,
\end{eqnarray}
\begin{eqnarray}
I_b(\xi) & = &\left\{\begin{array}{ll}
 \int_\xi^\infty K_{1/3}(x)\dif{x} & \gammadeux > 0\\
 \pi\sqrt{3} - \int_\xi^\infty \dif{x}F_{1/3}(x,\gammadeux)  & \gammadeux < 0
\end{array}\right. ,
\end{eqnarray}
\begin{eqnarray}
I_c(\xi) & = &\left\{\begin{array}{ll}
\int_\xi^\infty K_{5/3}(x)\dif{x}  -  K_{2/3}(\xi)  & \gammadeux > 0 \\
 \pi\sqrt{3} -  \int_\xi^\infty \dif{x}F_{1/3}(x,\gammadeux) - & \gammadeux < 0 \\
 F_{2/3}(\xi,\gammadeux)  
\end{array} \right. ,
\end{eqnarray}
\begin{eqnarray}
I_d(\xi) & = &  -4\pi\sqrt{3}\sqrt[3]{\frac{4}{3\xi}} \int_{-\infty}^\infty \dif{x} \airy(x^2 + c) \airy'(x^2 + c) ,
\end{eqnarray}
\begin{eqnarray}
I_e(\xi) & = & \frac{10}{3\xi}F_{1/3}\left(\xi,\gammadeux\right) + \\
 & & \left\{
\begin{array}{ll}
 \int_{\xi}^{+\infty}\dif{x}F_{1/3}(x,\gammadeux) - F_{2/3}(\xi,\gammadeux) & \gammadeux > 0 \\
\pi\sqrt{3} - \int_{\xi}^{+\infty}\dif{x}F_{1/3}(x,\gammadeux) - F_{2/3}(\xi,\gammadeux) & \gammadeux < 0
\end{array}\right. , \nonumber
\end{eqnarray}
\begin{eqnarray}
\label{eqIf}
I_f(\xi) & = &  \frac{2}{3\xi} F_{1/3}\left(\xi,\gammadeux\right)  + \\
& & \left\{
\begin{array}{ll}
 \int_{\xi}^{+\infty}\dif{x}F_{1/3}(x,\gammadeux) - F_{2/3}(\xi,\gammadeux)  & \gammadeux > 0 \\
\pi\sqrt{3} - \int_{\xi}^{+\infty}\dif{x}F_{1/3}(x,\gammadeux) - F_{2/3}(\xi,\gammadeux) & \gammadeux < 0
\end{array}\right.  \nonumber.
\end{eqnarray}
Among these, only $I_d$ could not be turned into a more convenient analytical form. Therefore we give here only its raw expression. The $F$ functions are defined as follow
\begin{equation}
\label{eqFfunc}
\begin{array}{l}
F_{1/3}(\xi, s) =  \left\{\begin{array}{ll}
K_{1/3}(\xi) &, s > 0 \\
\frac{\pi}{\sqrt{3}}\left( J_{1/3}(\xi) + J_{-1/3}(\xi)\right) &, s < 0
\end{array}\right., \\
F_{2/3}(\xi,s)  = \left\{\begin{array}{ll}
K_{2/3}(\xi) &, s > 0 \\
\frac{\pi}{\sqrt{3}}\left( J_{2/3}(\xi) - J_{-2/3}(\xi)\right) &, s<0
\end{array}\right. . 
\end{array}
\end{equation}

Performing replacements \eqref{eqrep1}-\eqref{eqrep6} we obtain the spectra per unit pulsation
\begin{widetext}
\begin{eqnarray}
\label{eqI1}
\deriv{I_{nn}^{\parallel}}{\omega}  & = & \frac{1}{2\pi\Omega}\frac{e^2\Omega \omega}{\gammatild^2\sqrt{3}4\pi\epsilon_0 c} \left[I_a(\xi) + \bcsb\omec\frac{\gamma^2}{\gammadeux}(n+1)\left( I_a(\xi) - \frac{4}{3} \frac{\omcrittilde}{\omega}I_d(\xi) \right) +  \omec n^2 I_c(\xi)\right], \\
\deriv{I_{nn}^{\perp}}{\omega} &  = &  \frac{1}{2\pi\Omega}\frac{e^2\Omega \omega}{\gammatild^2\sqrt{3}4\pi\epsilon_0 c} \left[I_c(\xi) + \bcsb\omec\frac{\gamma^2}{\gammadeux}(n+1)I_c(\xi) + \omec n^2 I_a(\xi) \right], \\
\deriv{I_{nn-1}^{\parallel}}{\omega} & = &  
 \frac{1}{2\pi\Omega}\frac{e^2\Omega \omega}{\gammatild^2\sqrt{3}4\pi\epsilon_0 c} \left[\frac{B}{B_c}n\frac{\gammatild^2}{\gamma^2}I_b(\xi) + \ome n \left(2\frac{\gammatild^2}{\gammadeux} I_b(\xi) + I_c(\xi)\right) + \right. \\
 & & \left. \bcsb\omec\frac{n}{2}\left(\frac{\gamma^2}{4\gammatild^2}I_e(\xi) + 2\frac{\gamma^2}{\gammatild^2}I_b(\xi) + 2\frac{\gamma^2}{\gammadeux}I_c(\xi) + \frac{3}{4}\frac{\gamma^2}{\gammatild^2}I_f(\xi) \right) +  \omec\frac{n^2}{4}\left(4\frac{\gammatild^2}{\gammadeux}I_b(\xi) - \frac{n-1}{n}I_c(\xi)\right) \right], \nonumber \\
\deriv{I_{nn-1}^{\perp}}{\omega} & = &  \frac{1}{2\pi\Omega}\frac{e^2\Omega \omega}{\gammatild^2\sqrt{3}4\pi\epsilon_0 c} \left[\frac{B}{B_c}n\frac{\gammatild^2}{\gamma^2}I_b(\xi) - \ome n I_c(\xi) + \bcsb\omec\frac{n}{8}\frac{\gamma^2}{\gammatild^2}\left(I_e(\xi) +  3I_f(\xi)\right) + \right. \nonumber \\
 & & \left. \omec\frac{n^2}{4}\left(4\frac{\gammatild^2}{\gammadeux}I_b(\xi) + \frac{n-1}{n}I_c(\xi)\right)\right], \\
\deriv{I_{nn-2}^{\parallel}}{\omega} & = &  \frac{1}{2\pi\Omega}\frac{e^2\Omega \omega}{\gammatild^2\sqrt{3}4\pi\epsilon_0 c} \left[\omec\frac{n(n-1)}{4}\left(I_a(\xi) + I_c(\xi)\right)\right], \\
\label{eqI4}
\deriv{I_{nn-2}^{\perp}}{\omega} & = &  \frac{1}{2\pi\Omega}\frac{e^2\Omega \omega}{\gammatild^2\sqrt{3}4\pi\epsilon_0 c} \left[\omec\frac{n(n-1)}{4}I_a(\xi)\right].
\end{eqnarray}
\end{widetext}

To have an estimate of the position of the peak of these spectra, following the arguments of the two previous sections one can take the critical pulsation without quantum correction for the $n\rightarrow n$ transitions, that is 
\begin{equation}
\label{eqomegac}
\omega_c = \Omega\frac{\gamma^2}{1 + 2nB/B_c}.
\end{equation} 
However, the other transitions cannot be treated exactly with the same arguments as in section \ref{seccourbure}, \eqref{eqomcritcourbure} owing to the fact that the factor $\gammadeux$ becomes infinite at a pulsation 
\begin{equation}
\label{eqomegapdown}
\omega_0 = \frac{E}{\hbar}\frac{2\Delta n B/B_c}{1 + 2nB/B_c} + \gdo{\frac{\hbar\omega}{E}}.
\end{equation}
If we restrict our reasoning to positive $\gammadeux$, or equivalently $\omega > \omega_0$, one can then show that the position of the peak of the spectra given above can be estimated to be 
\begin{equation}
\label{eqpeakpulsation}
\omega_p \sim \max(\omega_c,\omega_0).
\end{equation}

\section{Discussion and conclusion \label{secconclusion}}
 \begin{figure*}
 \includegraphics[width = 1\textwidth]{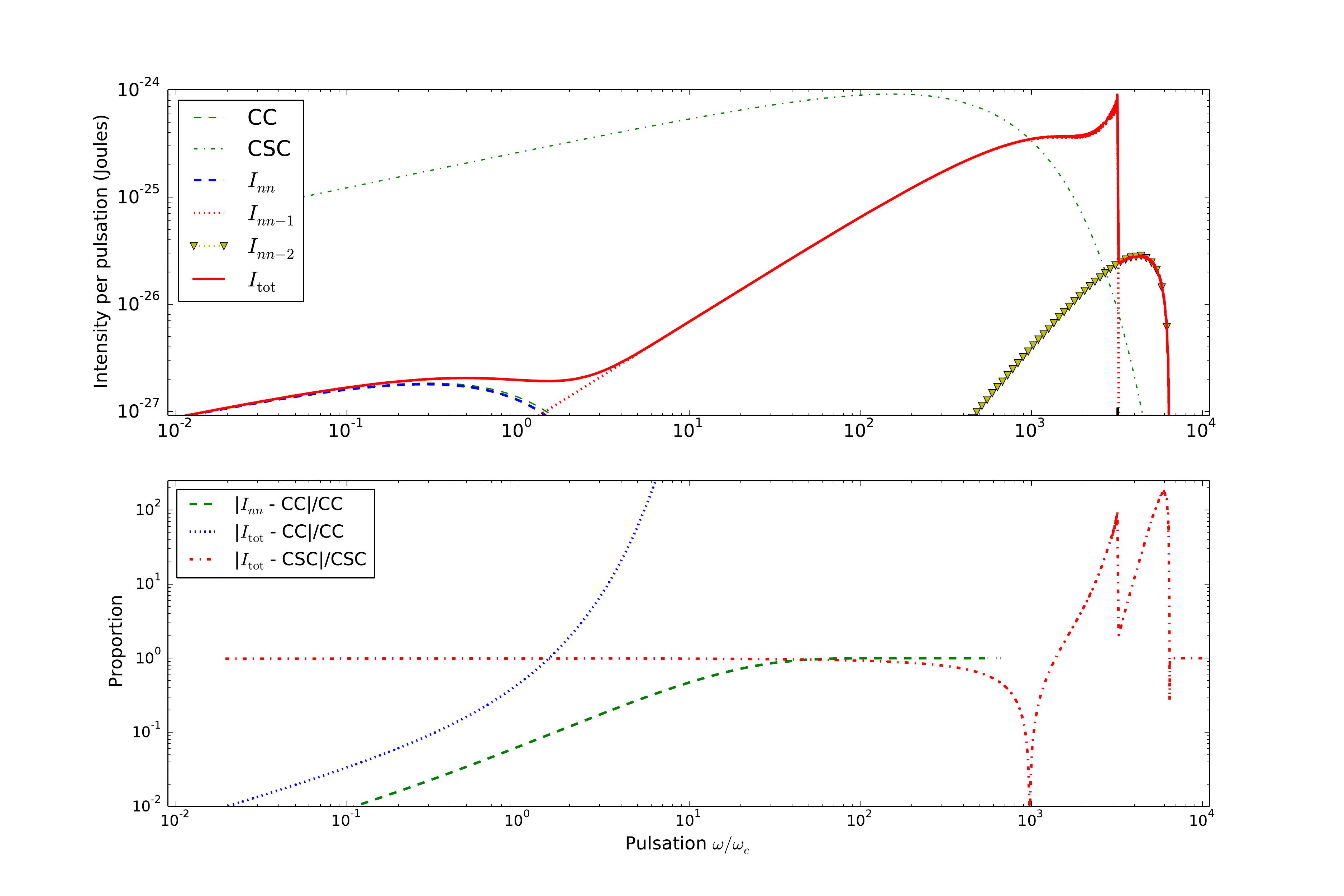}
 \caption{\label{figspectre} Upper panel: Intensities radiated by an electron following a magnetic field of radius of curvature $4\dix{4}$m, intensity $10^6$ Teslas, at a Lorentz factor $\gamma = 10^5$, on a perpendicular level $n=100$. For comparison, classical curvature (CC) and classical synchro-curvature (CSC, formula of \cite{vigano_compact_2014}) radiation are plotted in dashed green and dot-dashed green respectively. The thicker lines are showing plots of formulas \eqref{eqI1}-\eqref{eqI4} summed over photon polarizations, respectively  the curvature component $I_{nn}$ in dashed blue, the first downward component $I_{nn-1}$ in dotted red and the second downward component in dotted yellow with trident markers in yellow. The sum of this three components $I_{\mathrm{tot}}$is plotted in plain red. Abscissa are scaled by the pulsation $\omega_c$ \eqref{eqomegac} and the thick ticks on lower axes show the position of the peak pulsation $\omega_p$ \eqref{eqomegapdown} of the downward components. 
 The lower panel shows the relative differences between the curvature component $I_{nn}$ and CC in dashed green (not represented on the full range because these components are getting numerically too small at high pulsations), the sum of all components $I_{\mathrm{tot}}$ and CC in dotted blue, $I_{\mathrm{tot}}$ and CSC in dot-dashed red.  
  One sees that, in this case, the three peaks due to the curvature, first downward component and second downward component are distinct in the total spectrum, which corresponds well to CC at low pulsations and bridges the gap to CSC at high pulsations.  However, it should be noted that the difference between the total spectrum and CSC is roughly around $100\%$ of CSC everywhere.}
 \end{figure*}
	In section \ref{seccourbure} we showed that classical curvature radiation can be derived from first principles of quantum electrodynamics in a self-consistent manner within the ultra-relativistic approximation. Indeed, the usual derivation of curvature radiation assumes the limit of an unphysical trajectory, as mentioned in the introduction of the present paper and in \cite{voisin_curvature_sf2a}. Curvature radiation then results from transitions between states of different longitudinal quantum numbers $\lpara$ but both in the ground perpendicular level. The assumed ultra-relativistic regime allows us to consider $\lpara$ as a continuous variable and obtain a continuous spectrum. Perpendicular levels are the quantum analogues of classical rotation around the magnetic field. In the perpendicular ground level, or perpendicular fundamental, we showed in paper 1 that although orbital angular momentum around the field line is null, the particle is maintained on the field line through spin-magnetic-field interaction. Therefore, curvature radiation understood as the radiation of a particle following a magnetic-field line without "turning" around it should be seen as a purely quantum phenomenon. However, this is not enough to obtain the classical result: one has to consider that the particle wave-function is infinitely confined on the magnetic-field line, which is equivalently achieved by assuming $\hbar \rightarrow 0$, obviously the classical limit, or that the magnetic field intensity $B\rightarrow \infty$ in \eqref{eqphasescale}, and to neglect the quantum recoil effect in \eqref{eqdl00} by assuming that the emitted photon energy $\hbar \omega \ll E$, where $E$ is the energy of the radiating particle .
	
	In section \ref{secgeneralcase}, we consider the general case of synchro-curvature radiation in the regime of very low pitch angle, so low that the perpendicular energy of the particle must be quantified. This is, to our knowledge, the first time such derivation is made. Therefore, the radiation becomes the sum of continuous transitions of $\lpara$ and discrete transitions between perpendicular levels labeled by the integer $n$. Moreover, we take into account deconfinement and quantum recoil effects up to second order. We show that in the ultra-relativistic regime, transitions involving a change of perpendicular quantum number are significant only for $n \rightarrow  n- 1$ and $n\rightarrow n-2$ with a decreasing importance as the jump is larger. Transitions $n \rightarrow n$ are the generalization of curvature radiation on an arbitrary level $n$ from which they differ by an effective Lorentz factor \eqref{eqgammatild} and an amplified proportional weight of deconfinement terms (because proportional to $n$ or $n^2$). The two other transitions can be considered as the synchrotron part of synchro-curvature radiation. 
	
	At leading order, $n\rightarrow n$ transitions have the same polarization as the classical curvature radiation, $n \rightarrow  n- 1$ transitions are not polarized at all, and  $n \rightarrow  n- 2$ has a ratio between parallel and perpendicular polarization of $1 + I_c(\xi)/ I_a(\xi)$.

	It is out of the scope of this paper to proceed to a general exploration of the spectra generated by our final formulae \eqref{eqI1}-\eqref{eqI4} depending on  magnetic field $B/B_c$,  curvature radius $\rho$, Lorentz factor $\gamma$ and perpendicular level $n$. However we show in figure \ref{figspectre} a case with parameters compatible with a polar cap of recycled millisecond pulsar \citep{arons_pulsar_2009}, $B = 10^6$ Teslas, $\rho = 4\dix{4}$ meters, a moderate Lorentz factor of $10^5$, and a perpendicular level $n = 100$. These parameters fall within our approxmations given in \eqref{eqapproxs} and paper 1 equation 19.   On the upper panel of figure \ref{figspectre} we plot the curvature component $I_{nn}$ (to make notations lighter we remove here the $\frac{\mathrm{d}}{\mathrm{d}\omega}$) in dashed blue, $I_{nn-1}$ in dotted  red and $I_{nn-2}$ in dotted down-triangle	yellow. In order to compare we also plotted classical curvature radiation (CC) in dashed green and classical synchro-curvature radiation (CSC) in dot-dashed green. The pitch angle $\alpha$ is related to $n$ by 
\begin{equation}
	\alpha = \frac{\sqrt{2nB/B_c}}{\gamma},
\end{equation}
and here $\alpha \simeq 2\dix{-6}$. This value is quite easily reached in simulations of motion of an electron with classical-synchro-curvature-radiation losses in pulsar-like magnetic fields in \cite{vigano_compact_2014} or \cite{kelner_synchro-curvature_2015}. 

	 If one neglects radiation losses, or more physically that the particle remains for a while at levels around $n \sim 100$, then one can compare the sum of the intensities of the three above mentioned transitions $I_\mathrm{tot} = I_{nn} + I_{nn-1} + I_{nn-2}$ (figure \ref{figspectre}, upper panel) with the intensity of the classical curvature radiation (figure \ref{figspectre}, lower panel)  and with the intensity of the classical synchro-curvature radiation (figure \ref{figspectre}, lower panel) .

	 Until the peak of CC radiation,  $I_{nn}$ and CC are very close with a difference of a few percents and up to 10 percents, after which the difference mostly due to deconfinement terms (that grows with photon energy) reaches more than 100\% at high energies.  One would obtain a similar deviation in the fundamental curvature regime, $n=0$, but with a slightly higher Lorentz factor. 
	 
	 Transitions to lower perpendicular levels become important at high energies, taking over the vanishing $I_{nn}$ component in $I_\mathrm{tot}$ they get quite close to the high-energy part of the CSC spectrum. Slight wiggles on the ascending parts of spectra $I_{nn-1}$ and $I_{nn-2}$ make the line a little bit thicker on this graph around $\omega_0$ (thick black tick) and are due to the fact that $\gammadeux < 0$ (see \eqref{eqgammadeux}) at low photon pulsations and therefore these spectra are expressed by oscillatory Bessel functions in virtue of \eqref{eqIa}-\eqref{eqIf} below their peak pulsations (see discussion around \eqref{eqpeakpulsation}). In particular, it is responsible in the present case for the very sharp peak and cut-off of $I_{nn-1}$. Spectrum  $I_{nn-2}$ takes over just above $\omega_0$ and is responsible for the last maximum.
		
		 As a result, the total intensity $I_\mathrm{tot}$ is very close to CC radiation at low photon energies and becomes comparatively closer to CSC radiation at the highest energies. Although we see on the lower panel that the CSC spectrum is quasi-always $\sim 100\%$ or more more intense than  $I_\mathrm{tot}$, this agrees with the general tendency in the classical theory of synchro-curvature radiation to show broader spectra at high energies compared to curvature radiation while tending to the curvature spectrum at lower energies, see e.g. \cite{kelner_synchro-curvature_2015}. We also notice that this transition of behavior between quasi-curvature and synchro-curvature is much sharper in the quantum theory in the case of figure \ref{figspectre}. The sharpness of this transition  depends on the difference between $\omega_c$ and $\omega_0$: if $\omega_0 \gg \omega_c$ as is the case on figure \ref{figspectre} the downward components have more "time" to grow before they cut off, on the contrary if $\omega_0 < \omega_c$ or $\omega_0 \sim \omega_c$ the transition is much smoother or even insignificant and the spectrum resembles closely the classical curvature spectrum CC. 
		 
		 More generally, it comes out of equations \eqref{eqI1}-\eqref{eqI4} that the $n-1$ and $n-2$ components are increasing with the intensity of the magnetic field and with the perpendicular level $n$.  The Lorentz factor has a significant impact on the relative importance of the deconfinement terms since their relative importance to the main term grows like  $(\hbar\omcrit / E)^p \sim \gamma^{2p} $ where $p = 1,2$. In the case of terms going like $\propto B_c/B$, this can even lead them to become dominant at low magnetic field and high Lorentz factors. However, in this case one falls under the limitation of \eqref{eqapproxs} and our approximation starts to fail, needing computation of higher order terms.

		 
		 It is to be noticed that perpendicular upward transitions, from $n-1$ and $n-2$ to $n$ are also possible. As mentioned, the only difference between upward and downward transitions is in the effective Lorentz factor $\eqref{eqphasescale2}$. The  probability of upward transition is generally lower than the downward transitions because the effective Lorentz factor is lower. However, for very high Lorentz factors this difference becomes smaller. Because of the necessity of high Lorentz factors, the range of parameters where significant upwards rates can be computed safely is quite narrow ( see \eqref{eqphasescale2} and approximation 19 in paper 1). In the case of figure \ref{figspectre}, the upward spectra are not represented because they are numerically 0.  However, we can speculate on other configurations. First we can speculate beyond our approximations: our scheme remains convergent even outside the validity region, the results keep the same qualitative behavior as shown above, and approximation 19 of paper 1 is regularly overcome in classical calculations (see paper 1). For example, an electron with Lorentz factor of $6.3\dix{6}$ (reasonable in a pulsar magnetosphere gap) on the perpendicular level $n=100$ with a magnetic field of $10^6$ Teslas and a radius of curvature of $10^4$ m yields in this formalism  a ratio of $0.6$ between the first upward and first downward components. This last example suggests that the decay to the perpendicular fundamental may be slow and not monotonous if the Lorentz factor of the particle is high enough, and that a computation of the total radiated spectrum may need to take into account the random perpendicular jumps along the trajectory. This would especially be important due to the smallness of neutron-star magnetospheres.
		 
		 The particular case where we deal with a jump between the perpendicular fundamental and the first excited level can be also seen as the lowest spin flip transition possible, in the sense that the perpendicular fundamental is the only state having a non-degenerate spin state and the only way to flip the spin is therefore to go to the first level (see paper 1). This is what we called spin-flip curvature radiation in a preliminary work \cite{voisin_curvature_gamma2016}.

\appendix

\section{Integration of squared Airy integrals \label{apsquaredairy}}

Here we compute different expressions that differ slightly. We therefore detail the first case and then proceed faster for the others.

We use the functions $F_{1/3}$ and $F_{2/3}$ defined in \eqref{eqAiF}. We here give an alternative definition that will be useful in the developments of this appendix
\begin{equation}
\label{eqFintegrale}
\begin{array}{l}
F_{1/3}(\xi) = \sqrt{3} \int_0^{+\infty} \dif{x}\cos\left(\frac{3}{2}\xi\left(sx + \frac{x^3}{3}\right)\right) ,\\
F_{2/3}(\xi) =\sqrt{3} \int_0^{+\infty} \dif{x}x\sin\left(\frac{3}{2}\xi\left(sx + \frac{x^3}{3}\right)\right),
\end{array}
\end{equation}
with $s \in [-1,0,1]$. We recall their definition from \eqref{eqFfunc}
\begin{equation}
\begin{array}{l}
F_{1/3}(\xi, s) =  \left\{\begin{array}{ll}
K_{1/3}(\xi) &, s > 0 \\
\frac{1}{3^{2/3}\Gamma\left(\frac{2}{3}\right)} &, s = 0 \\
\frac{\pi}{\sqrt{3}}\left( J_{1/3}(\xi) + J_{-1/3}(\xi)\right) &, s < 0
\end{array}\right., \\
F_{2/3}(\xi,s)  = \left\{\begin{array}{ll}
K_{2/3}(\xi) &, s > 0 \\
-\frac{1}{3^{1/3}\Gamma\left(\frac{1}{3}\right)} &, s = 0 \\
\frac{\pi}{\sqrt{3}}\left( J_{2/3}(\xi) - J_{-2/3}(\xi)\right) &, s<0
\end{array}\right. . 
\end{array}
\end{equation}
They are related to the Airy function and its derivative by 
\begin{eqnarray}
\label{eqAiF}
 F_{1/3}(\xi, s) = \pi\sqrt{\frac{3}{\abs{x}}}\airy(x), \\
 \label{eqAiprimeF}
 F_{2/3}(\xi, s) = -\pi\frac{\sqrt{3}}{x}\airy'(x), 
\end{eqnarray}
where $x = \mathrm{sign}(s)\left(\frac{3}{2}\xi\right)^{2/3}$.

We also frequently use the following integrals 
\begin{equation}
\label{eqxnexp}
\int_{-\infty}^\infty \dif{\tau} \tau^n \exp\left[i a \tau^2\right] =  \frac{\Gamma\left(\frac{n+1}{2}\right)}{\abs{a}^{\frac{n+1}{2}}} e^{i\left(\frac{\pi}{4}-\frac{\arg(a)}{2}\right)(n+1)},
\end{equation}
where $a$ is a complex with argument $0< \arg(a) < \pi$ and $n$ a positive integer.  For practical purposes we give particular value of the $\Gamma$ function \cite{olver_nist_2010}
\begin{equation}
\Gamma\left(\frac{1}{2}\right) = \sqrt{\pi}, \Gamma\left(1\right) = 1,\Gamma\left(\frac{3}{2}\right) =  \frac{\sqrt{\pi}}{2}, \Gamma\left(\frac{5}{2}\right) =  \frac{3\sqrt{\pi}}{4}.
\end{equation}

\subsection{First case \label{firstcase} }
We want to compute the following expression, where $c$ is a constant 

\begin{equation}
\label{eqIAiry}
I_a = \frac{\sqrt{3}}{\pi\abs{c}}\int_{-\infty}^\infty \dif{x} \abs{\int_{-\infty}^\infty \dif{\tau} \tau \exp\left[\imath \left((c + x^2)\tau + \frac{\tau^3}{3}\right) \right] }^2.
\end{equation}

The present derivation is directly inspired by that of  \cite{cheng_general_1996}, however correcting for a mistake that we point out below.

Since 
\begin{equation}
A_i(y) = \frac{1}{2\pi}\int_{-\infty}^\infty \dif{\tau} \exp\left[\imath \left(y\tau + \frac{\tau^3}{3}\right) \right],
\end{equation}

one remarks that 
\begin{equation}
	I_a = \frac{\sqrt{3}}{\pi\abs{c}}\int_{-\infty}^\infty \dif{x} \abs{ 2\pi A_i'(c+x^2)}^2,
\end{equation}

where $A_i'$ is the derivative of the Airy function as defined in \cite{olver_nist_2010}. 


We will seek to evaluate $I$ through its integral formulation \ref{eqIAiry}. Developing the squared Airy integral we get 
\begin{equation}
\begin{array}{ll}
 I_a = & \frac{\sqrt{3}}{\pi\abs{c}}\int_{-\infty}^\infty \dif{x} \int_{-\infty}^\infty \dif{\tau_1} \int_{-\infty}^\infty \dif{\tau_2} \tau_1\tau_2 \\
 & \exp\left[\imath (\tau_1-\tau_2)\left((c + x^2) + \frac{1}{3}\left(\tau_1^2 + \tau_1\tau_2 + \tau_2^2\right)\right) \right] 
 \end{array}
\end{equation}

In order to "separate" as much as possible the integrals we introduce the following variables:
\begin{equation}
\label{eqvarchangeairyint}
\left(\tau_1, \tau_2 \right) \rightarrow \left(\tau_+ =  \frac{1}{2}\left(\tau_1 + \tau_2\right), \tau_- = \frac{1}{2}\left(\tau_1 - \tau_2\right) \right) 
\end{equation}

The Jacobian of this transformation is:

 \begin{equation}
 \abs{\pderiv{(\tau_1,\tau_2)}{(\tau_+,\tau_-)}} = \abs{\begin{matrix}
 1 & 1 \\
 1 & -1
 \end{matrix} } = 2
 \end{equation}

And we notice that:
\begin{eqnarray}
	\tau_1\tau_2  & = & \tau_+^2 - \tau_-^2 \\
	\tau_1^2 + \tau_1\tau_2 + \tau_2^2 & = & 3\tau_+^2 + \tau_-^2
\end{eqnarray}

Such that we get the form:

\begin{equation}
\begin{array}{ll}
	I_a =  & \frac{\sqrt{3}}{\pi\abs{c}}\int_{-\infty}^\infty \dif{x} \left\{ 2 \int_{-\infty}^\infty \dif{\tau_+}  \int_{-\infty}^\infty \dif{\tau_-}  \right.\\
	& \left. \exp\left[2\imath \tau_-\left((c + x^2)+ \frac{\tau_-^2}{3}\right) \right] \left(\tau_+^2 - \tau_-^2 \right) \right. \\
	& \left. \exp\left[2\imath\tau_-\tau_+^2\right] \right\}
	\end{array}
\end{equation}

Here one splits the computation in two integrals 
\begin{equation}
\begin{array}{ll}
	C  = & \int_{-\infty}^\infty \dif{x} \int_{-\infty}^\infty \dif{\tau_+} \int_{-\infty}^\infty \dif{\tau_-}  \\
	& \exp\left[2\imath\tau_-\tau_+^2\right] \exp\left[2\imath \tau_-\left((c + x^2)+ \frac{\tau_-^2}{3}\right) \right] \tau_-^2 \\
	D  = & \int_{-\infty}^\infty \dif{x}  \int_{-\infty}^\infty \dif{\tau_+} \tau_+^2\int_{-\infty}^\infty \dif{\tau_-}  \\
	& \exp\left[2\imath\tau_-\tau_+^2\right]  \exp\left[2\imath \tau_-\left((c + x^2)+ \frac{\tau_-^2}{3}\right) \right] 
\end{array}
\end{equation}
Such that:
\begin{equation}
I_a = 2\frac{\sqrt{3}}{\pi\abs{c}}(D - C)
\end{equation}

Here it would be nice to integrate over $\tau_+$ first since these integrals are of Gaussian type. However, the integrals cannot be swapped in $D$ without becoming divergent, as done in \cite{cheng_general_1996}. We will circumvent this problem by introducing a positive real parameter $\epsilon$ ,
\begin{equation}
\begin{array}{ll}
	D = & \lim_{\epsilon \rightarrow 0^+}  \\
	& \int_{-\infty}^\infty \dif{x}  \int_{-\infty}^\infty \dif{\tau_+} \tau_+^2  \int_{-\infty}^\infty \dif{\tau_-}\exp\left[2\imath(\tau_- + \imath \epsilon)\tau_+^2\right] \\
	&  \exp\left[2\imath (\tau_- + \imath \epsilon) \ \left((c + x^2)+ \frac{(\tau_- + \imath \epsilon)^2}{3}\right) \right] 
\end{array}
\end{equation}
which is allowed by the theorem of dominated convergence using for example the following hat function 
\begin{equation}
\begin{array}{l}
 g(\tau_+,\tau_-) = \\
  \left\{\begin{array}{l}
 \tau_+^2\left({max}\left(0,  \cos\left[ \tau_-\left((c + x^2)+ \tau_+^2 + \frac{\tau_-^2}{3}\right)\right]\right) + \right.\\
 \left. \imath \text{max}\left(0, \sin\left[ \tau_-\left((c + x^2)+ \tau_+^2 + \frac{\tau_-^2}{3}\right)\right]\right) \right) \\
 \abs{\tau_-}\tau_+^2\left({max}\left(0,  \cos\left[ \tau_-\left((c + x^2)+ \tau_+^2 + \frac{\tau_-^2}{3}\right)\right]\right) + \right.\\
 \left. \imath \text{max}\left(0, \sin\left[ \tau_-\left((c + x^2)+ \tau_+^2 + \frac{\tau_-^2}{3}\right)\right]\right) \right)
 \end{array}\right.
 \end{array}.
\end{equation}

Then we can first integrate over $\tau_+$ using \eqref{eqxnexp},

%

\begin{equation}
\begin{array}{ll}
C = & \int_{-\infty}^\infty \dif{x}   \int_{-\infty}^\infty \dif{\tau_-}  \sqrt{\frac{\pi}{2\abs{\tau_-}}} e^{\imath\frac{\pi}{4}s_{\tau_-}} \\
& \exp\left[2\imath \tau_-\left( (c + x^2) + \frac{\tau_-^2}{3}\right) \right]  \tau_-^2 , \\
D  =& \lim_{\epsilon \rightarrow 0^+} \int_{-\infty}^\infty \dif{x}  \int_{-\infty}^\infty \dif{\tau_-}  \sqrt{\frac{\pi}{2\abs{\tau_- + \imath \epsilon}}} e^{\imath\frac{\pi}{4}s_{\tau_-} + i \Delta_\epsilon}  \\
& \exp\left[2\imath(\tau_- + \imath \epsilon)\left( (c + x^2) + \frac{(\tau_- + \imath \epsilon)^2}{3}\right) \right] \frac{\imath}{4(\tau_- + \imath \epsilon)} 
\end{array},
\end{equation}

where $\Delta_\epsilon = -\frac{1}{2}\left(\text{arg}(x+ \imath \epsilon) - \text{arg}(x)\right)$.

	Summing over $x$,
\begin{equation}
\begin{array}{ll}
	C  = & \int_{-\infty}^\infty \dif{\tau_-}  \frac{\imath\pi}{2}\tau_-  \exp\left[2\imath \tau_-\left( c  + \frac{\tau_-^2}{3}\right) \right] \\
	D = & \lim_{\epsilon \rightarrow 0^+} \int_{-\infty}^\infty \dif{\tau_-}   \frac{-\pi}{8(\tau_- + \imath \epsilon)^2}  \\
	& \exp\left[2\imath(\tau_- + \imath \epsilon)\left( c + \frac{(\tau_- + \imath \epsilon)^2}{3}\right) \right] 
\end{array}
\end{equation}

Performing the following change of variable in $C$ 
\begin{equation}
 \tau_- \rightarrow y = \frac{2}{\sqrt[3]{4}}\tau_-,
\end{equation}
we recognize that $C$ is proportional to the derivative of the Airy integral with respect to $c' = \sqrt[3]{4}c$. Expressing it with a modified Bessel function according to \ref{eqAiprimeF} 

\begin{equation}
C = \left\{ \begin{array}{lc}
-\pi\frac{c}{\sqrt{3}} K_{2/3}\left(\xi\right) & c > 0\\
\pi^2\frac{c}{3}\left(J_{2/3}(\xi) - J_{-2/3}(\xi)\right)	& c < 0
\end{array}\right.
\end{equation}
where $ \xi = \frac{4}{3}c^{3/2}$ .

For $D$, we perform the following change of variable  
\begin{equation}
\label{eqvarchangeairyint2}
 \tau_- \rightarrow y = \frac{1}{\sqrt{\abs{c}}}\left(\tau_- + \imath \epsilon\right)
\end{equation}

Which means integrating in the complex plane on the line defined by $y = \frac{i\epsilon}{\sqrt{\abs{c}}}$. Taking again $\xi = \frac{4}{3}\abs{c}^{3/2}$ and $s_c = \mathrm{sign}(c)$ we write

\begin{equation}
\label{eqDIa}
	D  =  \frac{-\pi}{8 c^{1/2}}\lim_{\epsilon \rightarrow 0^+} \int_{y = \frac{i\epsilon}{\sqrt{\abs{c}}}} \dif{y}  \frac{1}{y^2}  \exp\left[\frac{3}{2}\xi\imath\left( s_c y + \frac{y^3}{3}\right) \right] .
\end{equation}

After integration by parts 

\begin{equation}
		D  =  \frac{-\imath \pi c}{ 4}\lim_{\epsilon \rightarrow 0^+} \int_{y = \frac{i\epsilon}{\sqrt{\abs{c}}}} \dif{y} \left(\frac{s_c}{y} +y \right)  \exp\left[\frac{3}{2}\xi\imath\left( y + \frac{y^3}{3}\right) \right] .
\end{equation}

Here we can swap again the integral and the limit except for the cosine part of the $1/y$ term. Indeed, if we go back to the $\tau_- = \sqrt{c}y - \imath\epsilon $ variable we see that 

\begin{equation}
\begin{array}{lcl}
p_\epsilon(\tau_-) & = & \frac{\cos\left[\frac{3}{2}\xi\left( y + \frac{y^3}{3}\right) \right]}{y} \\
& = & \frac{\cos\left[2\left( c(\tau_- + \imath\epsilon) + \frac{(\tau_- + \imath\epsilon)^3}{3}\right) \right]}{(\tau_- + \imath\epsilon)} \\
\end{array}.
\end{equation} 

We see that because of the pole in $\tau_- = 0$ it is impossible to find a hat function $g$ such that 
\begin{equation}
	\forall \epsilon> 0, \forall \tau_- \in \mathbb{R},  g(\tau_-) > \abs{p_\epsilon(\tau_-)},
\end{equation}
and therefore the swapping is forbidden. 

However, we may compute $\lim_{\epsilon\rightarrow 0} \int_{-\infty}^{+\infty}\dif{\tau_-}p_\epsilon(\tau_-)$ directly. Let's first write 

\begin{equation}
\begin{array}{ll}
	\forall \epsilon, L> 0,\int_{-\infty}^{+\infty}\dif{\tau_-}p_\epsilon(\tau_-) = & \\ \int_{-L}^{+L}\dif{\tau_-}p_\epsilon(\tau_-) + \underbrace{ \int_{\mathbb{R}\backslash [-L, L]} \dif{\tau_-}p_\epsilon(\tau_-) }_{(a)}.
	\end{array}
\end{equation}

The first term on the right-hand side can be written

\begin{equation}
\begin{array}{ll}
	 \int_{-L}^{+L}\dif{\tau_-}p_\epsilon(\tau_-) = &  \int_{-L}^{+L}\dif{\tau_-} \frac{\cos \left(2c\left( \tau_- + i\epsilon\right) \right)}{\tau_- + i\epsilon} \\
	 & + \circ\left( (L + \imath\epsilon)^5 \right).
	 \end{array},
\end{equation}
where the notation $\circ(x)$ is to be understood as $\circ(x) = xf(x)$ where $f$ is analytical and tends to $0$ as $x$ tends to $0$.

The first term on the right-hand side can be expressed as 

\begin{equation}
\begin{array}{ll}
  \int_{-L}^{+L}\dif{\tau_-} \frac{\cos \left(2c\left( \tau_- + i\epsilon\right) \right)}{\tau_- + i\epsilon} = &  \int_{-\infty}^{+\infty}\dif{\tau_-} \frac{\cos \left(2c\left( \tau_- + i\epsilon\right) \right)}{\tau_- + i\epsilon} - \\
  &    \underbrace{\int_{\mathbb{R}\backslash [-L, L]}\dif{\tau_-} \frac{\cos \left(2c\left( \tau_- + i\epsilon\right) \right)}{\tau_- + i\epsilon} }_{(b)}.
  \end{array}.
\end{equation}

The  first term on the right-hand side can be developed as 
\begin{equation}
\begin{array}{ll}
\int_{-\infty}^{+\infty}\dif{\tau_-} \frac{\cos \left(2c\left( \tau_- + i\epsilon\right) \right)}{\tau_- + i\epsilon}  = &  \cos\left(2ci\epsilon\right) \int_{-\infty}^{+\infty}\dif{\tau_-} \frac{\cos \left(2c\tau_- \right)}{\tau_- + i\epsilon}    -  \\ 
&\underbrace{ \sin\left(2ci\epsilon\right)\int_{-\infty}^{+\infty}\dif{\tau_-} \frac{\sin \left(2c \tau_- \right)}{\tau_- + i\epsilon} }_{(c)}
\end{array}.
\end{equation}

The integral in $(b)$ is a well-known integral \citep{gradshteyn_table_2000} given by
\begin{equation}
 \int_{-\infty}^{+\infty}\dif{\tau_-} \frac{\cos \left(2c \tau_-  \right)}{\tau_- + i\epsilon} = -\imath\pi e^{-2c\epsilon}.
\end{equation}

Now we can take the limit $\epsilon \rightarrow 0$. One can obviously swap integral and limit in $(a)$, $(b)$ and $(c)$. $(a)$ and $(b)$ cancels because the integrand is odd while $(c)$ cancels because of the sine prefactor. It follows that 

\begin{equation}
\label{eqcosineovery}
\lim_{\epsilon\rightarrow 0} \int_{-\infty}^{+\infty}\dif{\tau_-}p_\epsilon(\tau_-) = -\imath\pi +\circ(L^5).
\end{equation}

Since the left-hand side does not depend on $L$ it follows that $\circ(L^5)$ is a constant proportional to $L^5$, namely $0$.

For the other terms in $D$, we swap limit and integral. When $c > 0$ we use the following relations demonstrated by \cite{schwinger_classical_1949} ( \cite{schwinger_classical_1949} uses the definitions of \cite{watson_bessel_1966} for the Bessel functions while we use those, slightly different, of \cite{olver_nist_2010}. However one can show that the relations \ref{eqKschwi1} and \ref{eqKschwi2} are not affected by the change of convention.)

\begin{equation}
\begin{array}{ll}
\label{eqKschwi1}
\int_{0}^{+\infty} \dif{x} \frac{\sin\left(\frac{3}{2}\xi\left(x + \frac{x^3}{3}\right)\right)}{x}  = & 
\frac{\pi}{2} - \frac{1}{\sqrt{3}} \int_{\xi}^{+\infty} \dif{x} K_{1/3}(x), \\
\end{array}
\end{equation}
and
\begin{equation}
\begin{array}{ll}
\label{eqKschwi2}
\int_{0}^{+\infty} \dif{x} \left(\frac{1}{x} + 2x \right) \sin\left(\frac{3}{2}\xi\left(x + \frac{x^3}{3}\right)\right) = & \\
\frac{\pi}{2} + \frac{1}{\sqrt{3}} \int_{\xi}^{+\infty} \dif{x} K_{5/3}(x),
\end{array}
\end{equation}

to obtain
\begin{eqnarray}
D & = &  \frac{\pi c}{2}\left( - \frac{1}{\sqrt{3}}\int_\xi^\infty K_{1/3}(x)\dif{x}  + \frac{1}{\sqrt{3}} K_{2/3}(\xi) \right) \label{eqairyintD} \\
  & = & \frac{\pi c}{2}\left( \frac{1}{\sqrt{3}}\int_\xi^\infty K_{5/3}(x)\dif{x}  - \frac{1}{\sqrt{3}} K_{2/3}(\xi) \right).
\end{eqnarray}

Here, \cite{cheng_general_1996} find a result exactly three times larger. We successfully compared our results with direct numerical integrations. 

Finally when $c > 0$
\begin{equation}
 \label{eqairyint1}
I_a =\int_\xi^\infty K_{5/3}(x)\dif{x}   + K_{2/3}(\xi).
\end{equation}

The case $c < 0$ needs to demonstrate the equivalent of \eqref{eqKschwi1} and \eqref{eqKschwi2} when $c < 0$. The demonstration is similar to that of \cite{schwinger_classical_1949}. Let's first notice that 
\begin{equation}
\begin{array}{ll}
\label{eqF13moi1}
\deriv{}{\xi}\int_{-\infty}^{+\infty} \dif{x} \frac{\sin\left(\frac{3}{2}\xi\left(-x + \frac{x^3}{3}\right)\right)}{x}  = & \\
 \int_{-\infty}^{+\infty} \dif{x} \frac{3}{2}\left(-x + \frac{x^3}{3}\right)\cos\left(\frac{3}{2}\xi\left(-x + \frac{x^3}{3}\right)\right).
 \\
\end{array}
\end{equation}
In the right-hand side, one recognizes an exact primitive minus a cosine term. The exact primitive cancels for reason of parity and we are left with 

\begin{equation}
\begin{array}{ll}
\deriv{}{\xi}\int_{-\infty}^{+\infty} \dif{x} \frac{\sin\left(\frac{3}{2}\xi\left(-x + \frac{x^3}{3}\right)\right)}{x}  = & \\
 -\int_{-\infty}^{+\infty} \dif{x} \cos\left(\frac{3}{2}\xi\left(-x + \frac{x^3}{3}\right)\right),
 \\
\end{array}
\end{equation}
where the right-hand side identifies with the function $F_{1/3}(\xi)$ in \eqref{eqFintegrale}.  Noticing that 
\begin{equation}
\lim_{L \rightarrow\infty} \int_{-\infty}^{+\infty} \dif{x} \frac{\sin\left(\frac{3}{2}L\left(-x + \frac{x^3}{3}\right)\right)}{x} = -\pi,
\end{equation}

we obtain, 

\begin{equation}
\label{eqschwicnegatif}
\begin{array}{l}
\int_{0}^{+\infty} \dif{x} \frac{\sin\left(\frac{3}{2}\xi\left(-x + \frac{x^3}{3}\right)\right)}{x} = \\
 -\frac{\pi}{2} + \frac{\pi}{3}\int_\xi^{+\infty}\dif{x} \left(J_{1/3}(x) + J_{-1/3}(x)\right) .
 \end{array}
\end{equation}

Using this and \eqref{eqFintegrale} we obtain $D$ and $I_a$ in the case $c < 0$,

\begin{equation}
\begin{array}{ll}
D = & \frac{\pi^2}{2} \abs{c} \left(1 -  \frac{1}{3}\int_\xi^\infty \dif{x}\left(J_{1/3}(x) + J_{-1/3}(x)\right) -\right. \\
& \left. \frac{1}{3}\left(J_{2/3}(\xi) - J_{-2/3}(\xi)\right)\right),
\end{array} 
\end{equation}
and, using functions $F$ \eqref{eqAiF},
\begin{equation}
I_a =   \pi\sqrt{3} -  \int_\xi^\infty \dif{x}F_{1/3}(x) - 3F_{2/3}\xi). 
\end{equation}

\subsection{Second case}

We compute 

\begin{equation}
\label{eqIAiry2}
I_b = \frac{\sqrt{3}}{2\pi}\int_{-\infty}^\infty \dif{x} \abs{\int_{-\infty}^\infty \dif{\tau} \exp\left[\imath \left((c + x^2)\tau + \frac{\tau^3}{3}\right) \right] }^2.
\end{equation}

Performing the change of variables \ref{eqvarchangeairyint} we get:

\begin{equation}
\begin{array}{ll}
	I_b = & \frac{\sqrt{3}}{2\pi}\int_{-\infty}^\infty \dif{x} \left\{ 2 \int_{-\infty}^\infty \dif{\tau_+}  \int_{-\infty}^\infty \dif{\tau_-}  \right. \\
	& \left. \exp\left[2\imath \tau_-\left((c + x^2)+ \frac{\tau_-^2}{3}\right) \right] \exp\left[2\imath\tau_-\tau_+^2\right] \right\}.
	\end{array}
\end{equation}

Integrating over $\tau_+$ we obtain

\begin{equation}
\begin{array}{ll}
	I_b = & 2\frac{\sqrt{3}}{2\pi}\int_{-\infty}^\infty \dif{x}   \int_{-\infty}^\infty \dif{\tau_-}  \sqrt{\frac{\pi}{2\abs{\tau_-}}} e^{\imath\frac{\pi}{4}s_{\tau_-}} \\
	& \exp\left[2\imath \tau_-\left( (c + x^2) + \frac{\tau_-^2}{3}\right) \right]. \\
	\end{array}
\end{equation}

Here we need to be careful to deal with the singularity of the cosine term. Consequently, before swapping the integrals and integrating over $x$ one must perform the change of variables \eqref{eqvarchangeairyint2}, then take the limit of the cosine term using \eqref{eqcosineovery} and  ×compute the sine term using \eqref{eqKschwi1} if $c > 0$ or \eqref{eqschwicnegatif} if $c < 0$. One eventually obtains 


 \begin{equation}
 \label{eqairyint2}
	I_b =\left\{\begin{array}{ll}
		\int_\xi^\infty K_{1/3}(x)\dif{x}  &, c > 0 \\
		\pi\sqrt{3} - \int_\xi^\infty \dif{x}F_{1/3}(x) &, c < 0
		\end{array}	\right.
\end{equation}

\subsection{Third case $I_c$}

We compute 

\begin{equation}
\label{eqIAiry3}
I_c = \frac{\sqrt{3}}{\pi\abs{c}}\int_{-\infty}^\infty \dif{x} x^2 \abs{\int_{-\infty}^\infty \dif{\tau}  \exp\left[\imath \left((c + x^2)\tau + \frac{\tau^3}{3}\right) \right] }^2.
\end{equation}

Here it is enough to see that the $x^2$ factor yields exactly the same result as the $\tau_+$ factor in $D$. Therefore

\begin{equation}
 \label{eqairyint3}
\begin{array}{l}
I_c = 2D = \\
\left\{\begin{array}{ll}
\int_\xi^\infty K_{5/3}(x)\dif{x}  -  K_{2/3}(\xi)  & c > 0 \\
 \pi\sqrt{3} -  \int_\xi^\infty \dif{x}F_{1/3}(x) -F_{2/3}(\xi)  & c < 0.
\end{array} \right.
\end{array}.
\end{equation}
Remark that we put here only the expression using $K_{5/3}$ , but one could also express it as a function of $K_{1/3}$ as in equation \ref{eqairyintD}.

\subsection{Fourth case $I_d$}

We want to compute 
\begin{eqnarray}
\label{eqIAiry4}
I_d & =& \frac{\sqrt{3}}{\pi\sqrt{\abs{c}}}\int_{-\infty}^\infty \dif{x} \left(\int_{-\infty}^\infty \dif{\tau}  \exp\left[\imath \left((c + x^2)\tau + \frac{\tau^3}{3}\right) \right] \right. \nonumber\\
& & \left.\int_{-\infty}^\infty \dif{\tau}\tau  \exp\left[\imath \left((c + x^2)\tau + \frac{\tau^3}{3}\right) \right] \right).
\end{eqnarray}

However we could not find a way to obtain a complete analytical expression for this integral. One has to compute it numerically using the following equivalent formula 
\begin{equation}
I_d  =   -4\pi\sqrt{\frac{3}{\abs{c}}} \int_{-\infty}^\infty \dif{x} \airy(x^2 + c) \airy'(x^2 + c).
\end{equation}

\subsection{Fifth case $I_e$}
We compute 
\begin{equation}
\label{eqIAiry5}
I_e = \frac{4\sqrt{3}}{\pi c^2}\int_{-\infty}^\infty \dif{x} x^2 \abs{\int_{-\infty}^\infty \dif{\tau} \tau \exp\left[\imath \left((c + x^2)\tau + \frac{\tau^3}{3}\right) \right] }^2.
\end{equation}
Performing the change of variable \eqref{eqvarchangeairyint} we get that 
\begin{equation}
\label{eqIedc}
I_e = 2\frac{4\sqrt{3}}{\pi c^2}(D-C) 
\end{equation}
with 
\begin{eqnarray}
C &  = &  \int_{-\infty}^\infty \dif{x} x^2 \int_{-\infty}^\infty \dif{\tau_+} \int_{-\infty}^\infty \dif{\tau_-} \tau_-^2  \\
	& & \exp\left[2\imath\tau_-\tau_+^2\right] \exp\left[2\imath \tau_-\left((c + x^2)+ \frac{\tau_-^2}{3}\right) \right],  \nonumber\\
D & = & \int_{-\infty}^\infty \dif{x} x^2 \int_{-\infty}^\infty \dif{\tau_+} \tau_+^2\int_{-\infty}^\infty \dif{\tau_-}   \\
	& & \exp\left[2\imath\tau_-\tau_+^2\right]  \exp\left[2\imath \tau_-\left((c + x^2)+ \frac{\tau_-^2}{3}\right) \right]. \nonumber
\end{eqnarray}

Integrating $C$ is quite straightforward by using two times \eqref{eqxnexp}, once for $\tau_+$, once for $x$ . One is left with an Airy integral and 
\begin{equation}
C = \frac{-\pi^2}{4\sqrt[3]{2}}\airy\left(2^{2/3}c\right).
\end{equation}

For $D$, as for $I_a$ in section \ref{firstcase} integrals cannot be exchanged without obtaining a divergent integrand. To avoid this we apply the same recipe, that is we introduce a positive real parameter $\epsilon$ such that 
\begin{eqnarray}
D & = & \lim_{\epsilon \rightarrow 0}\int_{-\infty}^\infty \dif{x} x^2 \int_{-\infty}^\infty \dif{\tau_+}  \tau_+^2\int_{-\infty}^\infty \dif{\tau_-}  \\
	& & e^{2\imath\left(\tau_- + i\epsilon\right)\tau_+^2  }e^{2\imath \left(\tau_- + i\epsilon\right)\left((c + x^2)+ \frac{\left(\tau_- + i\epsilon\right)^2}{3}\right) }. \nonumber
\end{eqnarray}
It is possible to invert the the integrals and we perform integration over $\tau_+$ and $x$ using \eqref{eqxnexp}. Performing the change of variable \eqref{eqvarchangeairyint2}, we get
\begin{equation}
\label{eqDf}
D = \lim_{\epsilon \rightarrow 0} \frac{-i\pi}{32\abs{c}}\int_{y = \frac{i\epsilon}{\sqrt{\abs{c}}}} \dif{y} \frac{e^{i\frac{3}{2}\xi\left(y s_c + \frac{y^3}{3}\right)}}{ y^3}
\end{equation}
where as before $s_c = \mathrm{sign}(c) $ and $\xi =  \frac{4}{3}\abs{c}^{3/2}$. Performing an integration by part we have 
\begin{equation}
D = \lim_{\epsilon \rightarrow 0} \frac{\pi\sqrt{\abs{c}}}{32}\int_{y = \frac{i\epsilon}{\sqrt{\abs{c}}}} \dif{y}\left(\frac{s_c}{y^2} + 1\right)e^{i\frac{3}{2}\xi\left(y s_c + \frac{y^3}{3}\right)}.
\end{equation}
The second term corresponds to $F_{1/3}(\xi,s_c)$ by definition \eqref{eqAiF}. The first term is, up to a factor, the same integral as in \eqref{eqDIa}.

With $D$ and $C$ we use formula \eqref{eqIedc} and expressing $C$ with a $F_{1/3}$ function using \eqref{eqAiF}, we obtain

\begin{widetext}
\begin{eqnarray}
\label{eqDf2}
D & = & \frac{\pi\abs{c}^{1/2}}{16\sqrt{3}} \left[ F_{1/3}\left(\xi\right)  + 2  \abs{c}^{3/2}\left\{
\begin{array}{ll}
 \int_{\xi}^{+\infty}\dif{x}F_{1/3}(x) - F_{2/3}(\xi)  & c > 0 \\
\pi\sqrt{3} - \int_{\xi}^{+\infty}\dif{x}F_{1/3}(x) - F_{2/3}(\xi) & c < 0
\end{array}\right. \right], \\
I_e & = &  \frac{5}{2\abs{c}^{3/2}}F_{1/3}\left(\xi\right) +\left\{
\begin{array}{ll}
 \int_{\xi}^{+\infty}\dif{x}F_{1/3}(x) - F_{2/3}(\xi) & c > 0 \\
\pi\sqrt{3} - \int_{\xi}^{+\infty}\dif{x}F_{1/3}(x) - F_{2/3}(\xi) & c < 0
\end{array}\right. .
\end{eqnarray}\end{widetext}

\subsection{Sixth case $I_f$}
We compute 
\begin{eqnarray}
\label{eqIAiry6}
I_f & = & \frac{4}{\pi\sqrt{3}c^2}\int_{-\infty}^\infty \dif{x} x^4 \\
& & \abs{\int_{-\infty}^\infty \dif{\tau} \exp\left[\imath \left((c + x^2)\tau + \frac{\tau^3}{3}\right) \right] }^2 \nonumber.
\end{eqnarray}
Performing the change of variable \eqref{eqvarchangeairyint} we get that 
\begin{eqnarray}
I_f & = & 2\frac{4}{\pi\sqrt{3}c^2}\int_{-\infty}^\infty \dif{x} x^4 \int_{-\infty}^\infty \dif{\tau_+} \int_{-\infty}^\infty \dif{\tau_-}   \\
	& & \exp\left[2\imath\tau_-\tau_+^2\right]  \exp\left[2\imath \tau_-\left((c + x^2)+ \frac{\tau_-^2}{3}\right) \right]. \nonumber
\end{eqnarray}
Is is not possible to to exchange integration over $x$ with integration over $\tau_-$. We work around this by inserting a positive real parameter $\epsilon$ 
\begin{eqnarray}
I_f & = & \frac{8}{\pi\sqrt{3}c^2}\lim_{\epsilon\rightarrow 0} \int_{-\infty}^\infty \dif{x} x^4 \int_{-\infty}^\infty \dif{\tau_+} \int_{-\infty}^\infty \dif{\tau_-}   \\
	& & e^{\left[2\imath\left(\tau_- + i\epsilon\right)\tau_+^2\right]}  e^{\left[2\imath \left(\tau_- + i\epsilon\right)\left((c + x^2)+ \frac{\left(\tau_- + i\epsilon\right)^2}{3}\right) \right]} , \nonumber
\end{eqnarray}
and perform integrations over $\tau_+$ and $x$ using \eqref{eqxnexp}. We then perform the change of variable \eqref{eqvarchangeairyint2} to obtain
\begin{eqnarray}
I_f & = & \frac{24}{\pi\sqrt{3}c^2}\lim_{\epsilon\rightarrow 0} \frac{-i\pi}{32\abs{c}}\int_{-\infty}^\infty \dif{y}   \frac{ e^{\left[\frac{3}{2}\xi \imath \left(ys_c + \frac{y^3}{3}\right) \right]}}{y^3}  , \nonumber
\end{eqnarray}
where $s_c = \mathrm{sign}(c)$. 

Here we recognize in the limit integral \eqref{eqDf}, which value is given in \eqref{eqDf2}. Therefore the final result is 
\begin{eqnarray}
I_f & = & \frac{1}{2\abs{c}^{3/2}} F_{1/3}\left(\xi\right)  + \\
& & \left\{
\begin{array}{ll}
 \int_{\xi}^{+\infty}\dif{x}F_{1/3}(x) - F_{2/3}(\xi)  & c > 0 \\
\pi\sqrt{3} - \int_{\xi}^{+\infty}\dif{x}F_{1/3}(x) - F_{2/3}(\xi) & c < 0
\end{array}\right. . \nonumber
\end{eqnarray}
\bibliography{Paper1-states_of_the_electron}

\end{document}